\documentclass[final,a4paper,3p]{elsarticle}

\journal{arxiv.org}

\usepackage{algorithmic}             
\usepackage[lined,ruled,linesnumbered,commentsnumbered]{algorithm2e}
\DontPrintSemicolon
\usepackage{array}                   
\usepackage{amsmath,amssymb,amsfonts,amsthm}
\allowdisplaybreaks
\usepackage{dsfont,bm,bbm} 
\usepackage{relsize}
\usepackage{mathrsfs}
\usepackage{booktabs}
\usepackage{mathtools}
\usepackage[german,english]{babel}   
\usepackage{caption}                 
\usepackage[utf8]{inputenc}          
\usepackage[T1]{fontenc}
\usepackage{float}
\usepackage{etoolbox}
\usepackage{graphicx}                
\usepackage{lmodern}                 
\usepackage{microtype}    
\usepackage{lineno}
\usepackage{multirow}
\usepackage{calligra}
\usepackage{nicefrac}
\usepackage[usenames,dvipsnames]{xcolor}
\usepackage{tikz}
\usepackage{tikz-network}
\usepackage[caption=false]{subfig}   
\usepackage[deletedmarkup=xout,authormarkup=none]{changes}
\usepackage{url} 
\usepackage[colorlinks]{hyperref}
\usepackage{cleveref}
\crefformat{equation}{(#2#1#3)}
\AtBeginDocument{%
  \hypersetup{
    citecolor=Black,
    linkcolor=Black,   
    urlcolor=Violet,
	pdfauthor={Uribe et al.}}}

\ifpdf
\hypersetup{
	pdftitle={sparsity promoting priors},
	pdfauthor={Uribe et al.}
}
\fi

\makeatletter
\renewcommand{\@algocf@capt@plain}{above}
\makeatother

\makeatletter
\def\url@leostyle{%
\@ifundefined{selectfont}{\def\UrlFont{\sf}}{\def\UrlFont{\small\ttfamily}}}
\makeatother
\newcommand*\patchAmsMathEnvironmentForLineno[1]{%
  \expandafter\let\csname old#1\expandafter\endcsname\csname #1\endcsname
  \expandafter\let\csname oldend#1\expandafter\endcsname\csname end#1\endcsname
  \renewenvironment{#1}%
     {\linenomath\csname old#1\endcsname}%
     {\csname oldend#1\endcsname\endlinenomath}}%
\newcommand*\patchBothAmsMathEnvironmentsForLineno[1]{%
  \patchAmsMathEnvironmentForLineno{#1}%
  \patchAmsMathEnvironmentForLineno{#1*}}%
\AtBeginDocument{%
\patchBothAmsMathEnvironmentsForLineno{equation}%
\patchBothAmsMathEnvironmentsForLineno{align}%
\patchBothAmsMathEnvironmentsForLineno{flalign}%
\patchBothAmsMathEnvironmentsForLineno{alignat}%
\patchBothAmsMathEnvironmentsForLineno{gather}%
\patchBothAmsMathEnvironmentsForLineno{multline}%
}

\newcommand{\given}{\;\ifnum\currentgrouptype=16 \middle\fi|\;}
\newcommand{\suchthat}{\;\ifnum\currentgrouptype=16 \middle\fi|\;}
\newcommand{\norm}[1]{\left\Vert#1\right\Vert}

\newcommand{\dd}{\mathrm{d}}
\newcommand{\tran}{\mathsf{T}}

\newcommand{\ve}[1]{\boldsymbol{#1}}
\newcommand{\mat}[1]{\boldsymbol{#1}}

\newcommand{\invg}{\text{IG}}
\newcommand{\stud}{\text{t}}

\DeclarePairedDelimiter\ceil{\lceil}{\rceil}

\DeclareMathOperator*{\argmax}{arg\,max}
\DeclareMathOperator*{\argmin}{arg\,min}
\DeclareMathOperator{\diag}{diag}

\newtheorem{theorem}{Theorem}[section]
\newtheorem{remark}[theorem]{Remark}

\newcommand{\yq}[1]{{\color{black}#1}}
\graphicspath{{figures/}}

\begin{document}

\begin{frontmatter}	
\title{Horseshoe priors for edge-preserving linear Bayesian inversion}

\author[DTU]{Felipe Uribe\corref{cor1}} \ead{furca@dtu.dk}
\author[DTU]{Yiqiu Dong} 
\author[DTU]{Per Christian Hansen} 

\cortext[cor1]{Corresponding author.} 

\address[DTU]{Department of Applied Mathematics and Computer Science, Technical University of Denmark. Richard Petersens Plads, Building 324, 2800 Kgs.\ Lyngby, Denmark.}

\begin{abstract}
In many large-scale inverse problems, such as computed tomography and image deblurring, characterization of sharp edges in the solution is desired. Within the Bayesian approach to inverse problems, edge-preservation is often achieved using Markov random field priors based on heavy-tailed distributions. Another strategy, popular in statistics, is the application of hierarchical shrinkage priors. An advantage of this formulation lies in expressing the prior as a conditionally Gaussian distribution depending of \emph{global} and \emph{local} hyperparameters which are endowed with heavy-tailed hyperpriors. In this work, we revisit the shrinkage horseshoe prior and introduce its formulation for edge-preserving settings. We discuss a sampling framework based on the Gibbs sampler to solve the resulting hierarchical formulation of the Bayesian inverse problem. In particular, one of the conditional distributions is high-dimensional Gaussian, and the rest are derived in closed form by using a scale mixture representation of the heavy-tailed hyperpriors. Applications from imaging science show that our computational procedure is able to compute sharp edge-preserving posterior point estimates with reduced uncertainty.
\end{abstract}

\begin{keyword}
Bayesian inverse problems, Bayesian hierarchical modeling, edge-preserving estimation, horseshoe prior, Gibbs sampler.
\vspace*{5pt}

\noindent \emph{AMS}: 62F15, 65C05, 65R32, 65F22.
\end{keyword}
\end{frontmatter}

\section{Introduction}
The integration of data into computational models is fundamental when one is interested in recovering unknown parameters defining the model. This specifies an {inverse problem} where the goal is to discover the parameters that make the model closely match the observed data. Inverse problems are typically ill-posed and the stability mainly depends on the structure of the mathematical model, the dimension of the parameter space, and the scarcity and noisiness of the data. Regularization methods are commonly used to find the solution of inverse problems (see, e.g., \cite{hansen_2010}). These methods are deterministic and incorporate penalty functions as stabilization procedure. Another approach relies on Bayesian statistics (see, e.g., \cite{kaipio_and_somersalo_2005}). These techniques incorporate a probabilistic description of the model parameters that combines prior information with a likelihood function that accounts for the model and data. In this approach, the objective is to estimate a so-called posterior distribution of the model parameters. Closed-form expressions of the posterior can only be derived in some particular cases, and in general the posterior has to be estimated using sampling-based methods such as Markov chain Monte Carlo (MCMC) \cite{gamerman_and_lopes_2006,owen_2019}, or approximation methods such as variational inference \cite{ranganath_et_al_2014}, transport maps \cite{elmoselhy_and_marzouk_2012} and Laplace approximation \cite{schillings_et_al_2020,uribe_et_al_2021}.

The main advantage of the Bayesian formulation lies in the possibility of quantifying parameter uncertainty due to noise and model errors. However, the approach is generally limited to problems where the dimension of the parameter space is not prohibitively large. In large-scale scenarios, the unknown parameters are modeled as random fields discretized pointwise on a fine grid. This complicates the application of sampling methods for Bayesian inference since one requires the exploration of a high-dimensional parameter space\,---\,even in the case of the linear problems considered here. Another level of complexity arises when the solution of the inverse problem is non-smooth and the preservation of the edges or sharp features in the solution is necessary \cite{bruckstein_et_al_2009}. This especially occurs in imaging science such as X-ray computed tomography, \yq{image deblurring, segmentation and denoising}. In these cases edge-preservation is generally imposed via the prior probability distribution. These prior models can be grouped into four categories:
\begin{itemize}
\item[(i)] \emph{Heavy-tailed Markov random fields} defined on pairwise parameter increments or differences. The idea is to increase the probability of large jump events by imposing heavy-tailed distributions on the increments. Some examples include total variation prior \cite{lassas_and_siltanen_2004}, Laplace Markov random fields \cite{bardsley_2012,uribe_et_al_2021} and Cauchy Markov random fields \cite{suuronen_et_al_2022}.
\item[(ii)] \emph{Random fields with jump discontinuities}, including Besov space priors based on Haar wavelet expansions \cite{dashti_et_al_2012,lassas_et_al_2009}, Gaussian and compound Poisson process priors \cite{hosseini_2017}, and level-set priors that employ Gaussian random fields with different thresholds via predefined level sets \cite{dunlop_et_al_2017}.
\item[(iii)] \emph{Machine learning-based models}, including implicit models such as plug-and-play priors \cite{kamilov_et_al_2017} and Bayesian neural networks with Cauchy weights that promote edge-preservation \cite{li_et_al_2022}.
\item[(iv)] \emph{Shrinkage priors}, also known as global-local or component-wise priors, that aim at shrinking small values towards zero while leaving the larger ones unaffected. These prior models are hierarchical by nature and include the methods proposed in \cite{calvetti_et_al_2019,calvetti_et_al_2020b}, as well as multiple models used in the statistics community such as elastic net, spike-slab, horseshoe, discrete Gaussian mixtures, and others (see, e.g., \cite{polson_and_sokolov_2019,vanerp_et_al_2019} for reviews).
\end{itemize}

One of the advantages of shrinkage priors is that they are essentially defined by conditionally Gaussian distributions. \yq{Given one or more hyperparameters such as the variance, a target uncertain parameter endowed with a Gaussian distribution can have many nice properties \cite{vono_et_al_2022}}. Different shrinkage priors differ in the definition of the hierarchical structure and probabilistic models used to represent the hyperparameters. For example, the Lasso prior is based on a scale mixture representation of the Laplace distribution, and thus it defines an exponential distribution on the variance with a half-Cauchy distributed rate parameter \cite{park_and_casella_2008}. The elastic net prior uses a truncated-gamma distribution on the variance with scale defined by two additional parameters that are half-Cauchy distributed \cite{li_et_al_2010}. Another shrinkage model is the horseshoe prior, in which the variance is defined by two hyperparameters that are half-Cauchy distributed \cite{carvalho_et_al_2010}; one hyperparameter is a scalar that controls the global variability, while the other is spatially dependent and captures local variations.

Despite being common in statistics, shrinkage priors are \yq{not} yet fully applied in large-scale linear inverse problems that require edge-preserving solutions. In this paper, we formulate a Bayesian inference approach that targets this requirement. We focus on the horseshoe prior, where the target parameter is Gaussian-distributed conditioned on half-Cauchy-distributed global and local hyperparameters. Our idea is that when this type of prior is formulated for the parameter increments or differences, the local hyperparameter assists in the identification of large discontinuities. The Bayesian inverse problem becomes hierarchical, since hyperparameters associated with the prior and likelihood are also part of the inference process.

The main difficulty in this formulation is that the resulting hierarchical posterior is challenging to handle computationally due to the large-scale nature of the target parameter and the heavy-tailed distributions imposed on the hyperparameters. To alleviate the complexity of the sampling process, we use the scale mixture representation of the half-Student-t distribution to express the half-Cauchy distributions of the hyperparameters in terms of inverse gamma distributions. Despite this requires the addition of extra hyperparameters, the hierarchical posterior is given essentially by products of Gaussian and inverse gamma densities. Therefore, we can exploit the conjugacy in the hierarchies to define a Gibbs sampler for which the associated full conditional densities are derived in closed form, and hence they can be sampled directly without an MCMC algorithm. However, within the Gibbs sampler, the conditional distribution of the target parameter defines a linear Gaussian Bayesian inverse problem. Although this full conditional is Gaussian and \yq{has an analytical expression}{, it is prohibitively slow to simulate in large-scale applications. Hence, the task of sampling from this distribution is formulated as a least-squares problem that is solved efficiently via standard iterative methods such as CGLS \cite{bjorck_1996} with a preconditioner based on the prior precision matrix. The computational framework is tested on a one-dimensional deconvolution problem, as well as, two-dimensional applications of image deblurring and computed tomography (CT).

\paragraph{Related work}
Existing applications of shrinkage priors in the context of inverse problems include the following. Bayesian inverse problems in sparse signal and image processing are surveyed in \cite{mohammaddjafari_2012} where a variety of shrinkage priors are discussed and applied to multiple problems in imaging science. MCMC and approximation methods are also suggested in \cite{mohammaddjafari_2012} to solve the Bayesian inverse problem depending on the choice of the prior. Spike-and-slab priors are used in \cite{riis_andersen_et_al_2017} for linear Bayesian inverse problems subject to sparsity constraints. The posterior is simulated with an expectation propagation algorithm using different approximations of the precision matrix of the underlying Gaussian parameter to improve the efficiency in high-dimensional applications. So-called \emph{Gaussian hypermodels} are proposed in \cite{calvetti_et_al_2019}, which are defined as conditionally Gaussian priors with a single local variance hyperparameter endowed with a gamma distribution. An iterative alternating sequential algorithm is proposed that converges to the maximum a posteriori probability estimator of the parameters. Following this idea, in \cite{calvetti_et_al_2020b} a generalized gamma distribution is imposed on the local variances and propose two modifications to their iterative algorithm to exploit the global convexity ensured by gamma hyperpriors and the stronger sparsity promotion of the generalized gamma hyperpriors.

\paragraph{Our contributions}
The main contributions of the paper are highlighted as follows:
\begin{itemize}
	\item[(i)] We review the horseshoe prior for the representation of parameters that are sparse. 
	\item[(ii)] We introduce the horseshoe prior for the solution of linear Bayesian inverse problems that require edge-preservation and perform uncertainty quantification.
	\item[(iii)] We propose a Gibbs sampler to solve the resulting hierarchical formulation of the Bayesian inverse problem; the full conditional distributions are derived in closed form by exploiting conjugacy. 
	\item[(iv)] We demonstrate the performance of our method through numerical experiments on two inverse problems.
\end{itemize}

\paragraph{Structure of the paper}
This paper is organized as follows. In \cref{sec:fram}, we briefly introduce the mathematical framework for linear inverse problems and the Bayesian approach, and we present the standard horseshoe prior. We then derive its generalization for edge-preserving linear Bayesian inversion in \cref{sec:HSedge}. We show that a scale mixture representation can be applied to express a heavy-tailed half-Cauchy hyperprior in terms of two inverse gamma distributions. The computational framework is defined in \cref{sec:method}, where we develop a Gibbs sampling approach to solve the resulting hierarchical Bayesian inverse problem. Here, we use CGLS to sample the \yq{full} Gaussian conditional and we derive the remaining hyperparameter conditionals in \yq{closed form}. In \cref{sec:numexp} we illustrate our computational approach with imaging science applications, before concluding in \cref{sec:conclusions}.

\section{Preliminaries}\label{sec:fram}
We commence with the mathematical framework and introduce linear inverse problems and the Bayesian approach to find solutions to those, as well as the standard horseshoe prior.

\subsection{Linear inverse problems}
We consider the discrete inverse problem of estimating an unknown model parameter $\ve{x}\in \mathbb{R}^{d}$ using noisy \yq{observed} data $\ve{y}\in \mathbb{R}^{m}$, given a linear forward operator $\mat{A}\in\mathbb{R}^{m\times d}$ acting as a link between the parameters and data. Here, $d$ is the number of parameters and $m$ denotes the number of data points. The parameter $\ve{x}$ is spatially varying and exists in connection with a physical domain $D$, e.g., one-dimensional signals, two-dimensional images and three-dimensional objects. Using a suitable finite-dimensional representation, the dimension $d$ will depend on a discretization size $N$ imposed on the physical domain. For instance, in one- and two-dimensional settings $d=N$ and $d=N^2$ (assuming equal discretization in both directions), respectively.

In particular, a linear inverse problem estimates an $\ve{x}$ that approximates the ground truth $\ve{x}^{\rm true}$ in
\begin{equation}\label{eq:inv_prob}
    {\ve{y}} = \mat{A}\ve{x}^{\rm true} + \ve{e}\qquad\text{with}\quad \ve{e}\sim \mathcal{N}(\ve{0}, \sigma_{\mathrm{obs}}^2\mat{I}_m),
\end{equation}
where the additive measurement noise is modeled as a realization of an independent Gaussian random vector with variance $\sigma_{\mathrm{obs}}^2$, $\mat{I}_m$ is an identity matrix of size $m$, and $\mathcal{N}\left(\cdot,\cdot\right)$ denotes the multivariate Gaussian distribution depending on mean and covariance parameters. The linear inverse problem associated to \eqref{eq:inv_prob} is typically ill-posed and we employ the Bayesian statistical framework to study the influence of noise in the observed data.

\subsection{Bayesian inverse problems}
In the Bayesian approach to inverse problems \cite{kaipio_and_somersalo_2005}, we treat $\ve{x},\ve{y}$ and $\ve{e}$ as random variables and the solution is expressed as the probability distribution of $\ve{x}$ given an instance of the observed data $\ve{y}$. This allows both modeling the noise via its statistical properties and specifying prior information on the parameter, i.e., the form of solutions that are believed more likely.

The unknown parameter is modeled as a discretized random field represented as a random vector $\ve{X}$ taking values $\ve{x}\in\mathbb{R}^{d}$. We assume the distribution of $\ve{X}$ has a so-called \emph{prior} probability density $\pi_{\mathrm{pr}}(\ve{x})$. The \emph{likelihood function} is defined from a density $\pi_{\mathrm{data}}(\cdot\given \ve{x})$ with fixed argument equal to the observed data $\ve{y}$; then the likelihood becomes a function of $\ve{x}$ only. The Gaussian statistical model for the measurement noise assumed in \eqref{eq:inv_prob} gives
\begin{equation}\label{eq:like}
\pi_{\mathrm{data}}(\ve{y}\given\ve{x}) =\frac{1}{(2\pi)^{\nicefrac{m}{2}} \sigma_{\rm obs}^m} \exp\left(-\frac{1}{2\sigma_{\rm obs}^2}\norm{\ve{y}-\mat{A}\ve{x}}_2^2\right).
\end{equation}

Assuming that the measurement noise and model parameters are independent, the solution of the Bayesian inverse problem combines the prior and likelihood probability models into a so-called \emph{posterior} density, given by Bayes' Theorem as
\begin{equation}\label{eq:Bayes}
\pi_{\mathrm{pos}}\left(\ve{x}\given \ve{y}\right) = \frac{1}{Z}\, \pi_{\mathrm{data}}(\ve{y}\given\ve{x})\,\pi_{\mathrm{pr}}(\ve{x}),
\end{equation}
where $Z = \int_{\mathbb{R}^d} \pi_{\mathrm{data}}(\ve{y}\given\ve{x})\,\pi_{\mathrm{pr}}(\ve{x})\, \dd \ve{x}$ is the normalizing constant of the posterior.

In practice, point estimates of the posterior \eqref{eq:Bayes} are used to represent the solution of the inverse problem. The two classical choices are the \emph{maximum a posteriori} (MAP) and the \emph{posterior mean} (PM) estimators:
\begin{equation}\label{eq:MAP}
\ve{x}_{\mathrm{MAP}} = \argmax_{\ve{x}\in\mathbb{R}^d} {\pi_{\mathrm{pos}}(\ve{x}\given \ve{y})} \quad \text{ and } \quad \ve{x}_{\mathrm{PM}} = \int_{\mathbb{R}^d}\ve{x}\,\pi_{\mathrm{pos}}(\ve{x}\given \ve{y})\,\dd\ve{x}.
\end{equation}

The MAP estimator is computed via optimization techniques which is often an advantage, compared to the calculation of the PM estimator that requires a more involved high-dimensional integration. The estimation of the posterior mean and related summary statistics is often performed via Monte Carlo methods (see, e.g., \cite{gamerman_and_lopes_2006,owen_2019}). We point out that when the first two statistical moments of a random vector do not exist (as it is common for some heavy-tailed distribution models), location and scale characteristics of the distribution can still be summarized using for instance the \emph{posterior median} and \emph{median absolute deviation}, respectively \cite{tenorio_2017}. These are also a sensible option to describe random variables whose distribution is heavily skewed.

To find the posterior \eqref{eq:Bayes}, we will focus on the definition of a prior probability model that allows both, the preservation of potential sharp features in the unknown parameter and a tractable computation of the high-dimensional posterior distribution. Our model is based on the horseshoe prior \cite{carvalho_et_al_2009}, which we revisit next.

\subsection{Horseshoe prior}
When our \emph{a priori} knowledge is that $\ve{x}$ is sparse, it may  be convenient to apply a sparsity-inducing shrinkage prior (see, e.g., \cite{vanerp_et_al_2019}). A well-known continuous shrinkage model is the standard horseshoe \cite{carvalho_et_al_2009}, which is defined as the hierarchical prior:
\begin{equation}\label{eq:pr_HS}
\pi_{\mathrm{pr}}(\ve{x},\tau,\ve{\sigma}) = \pi_{\mathrm{pr}}(\ve{x}\given \tau,\ve{\sigma})\pi_{\mathrm{hpr}}(\tau)\pi_{\mathrm{hpr}}(\ve{\sigma}),
\end{equation}
where a zero-mean conditionally Gaussian prior is imposed on $\ve{x}$:
\begin{equation}\label{eq:pr_x}
\pi_{\mathrm{pr}}(\ve{x}\given\tau,\ve{\sigma}) = \frac{1}{(2\pi)^{\nicefrac{d}{2}} (\det \mat{\Sigma}(\tau,\ve{\sigma}))^{\nicefrac{1}{2}}} \exp\left(-\frac{1}{2}\ve{x}^\tran \mat{\Sigma}^{-1}(\tau,\ve{\sigma}) \,\ve{x}\right),\qquad  \mat{\Sigma}(\tau,\ve{\sigma})=\tau^2\diag(\sigma_1^2, \ldots, \sigma_d^2).
\end{equation}
Here, the prior covariance matrix
$\mat{\Sigma}(\tau,\ve{\sigma})\in\mathbb{R}^{d\times d}$ depends on the hyperparameters $\tau\in \mathbb{R}_{>0}$ controlling \emph{global shrinkage} and $\ve{\sigma}=[\sigma_1, \ldots, \sigma_d]^\tran\in\mathbb{R}^d_{>0}$ defining \emph{local shrinkage}. The global-local scheme is defined with  heavy-tailed distributions. In particular, the horseshoe model uses a half-Cauchy distribution for $\tau$ and an independent standard half-Cauchy distribution for $\ve{\sigma}$:
\begin{equation}\label{eq:hyperpr}
\pi_{\mathrm{hpr}}(\tau) \propto\frac{2}{\tau_0\left(1+\frac{\tau^2}{\tau_0^2}\right)} \quad\text{and}\quad \pi_{\mathrm{hpr}}(\ve{\sigma}) \propto \prod_{i=1}^d\frac{2}{1+\sigma_i^2} \qquad \text{with}~\tau,\sigma_i>0,
\end{equation}
where $\tau_0$ is a scale parameter. According to existing work, we can set (i) $\tau_0\approx\sigma_{\rm obs}$ \cite{carvalho_et_al_2009}; (ii) $\tau_0=(d_0/(d-d_0))\sigma_{\rm obs}$, where $d_0$ is the number of nonzero elements  \cite{piironen_and_vehtari_2017}; (iii) $\tau_0$ as a part of the inference, e.g., using a Jeffreys' hyperprior \cite{carvalho_et_al_2010}.

Another expression for the horseshoe prior is given by re-writing the density of $\sigma_i$ as $\pi_{\mathrm{hpr}}(\sigma_i) \propto 1/(1+\sigma_i^2)=\kappa_i$ for $i=1,\ldots,d$, where $\kappa_i$ is called a shrinkage parameter. The relation between $\sigma_i$ and $\kappa_i$ is bijective when $\sigma_i>0$, then its inverse is continuously differentiable and non-zero. Therefore, the hyperprior of each parameter $\kappa_i$ is found from the half-Cauchy hyperprior of $\sigma_i$ using a standard probabilistic transformation and we obtain
\begin{equation}\label{eq:kappa}
\pi_{\mathrm{hpr}}(\kappa_i) \propto \frac{1}{\sqrt{\kappa_i}\sqrt{1-\kappa_i}},
\end{equation}
which is proportional to a \emph{horseshoe-shaped} beta probability density with shape parameters equal to $1/2$, hence the name of the prior. Small values of $\kappa_i$ correspond to $\sigma_i\to\infty$ and produce almost no shrinkage, while values $\kappa_i\approx 1$ corresponding to $\sigma_i\to 0$ provide essentially full shrinkage, and hence $\kappa_i$ can be used to describe how many active or inactive variables are present in the model \cite{carvalho_et_al_2010}.

The horseshoe prior is hierarchical by definition and lacks of a proper analytical expression after marginalizing out the hyperparameters. However, lower and upper bounds for the one-dimensional horseshoe probability density are available \cite{carvalho_et_al_2010}:
\begin{equation}
\frac{1}{2\sqrt{2\pi^3}}\log\left(1+\frac{4}{x^2}\right)\leq \pi_{\mathrm{pr}}(x) \leq \frac{1}{\sqrt{2\pi^3}}\log\left(1+\frac{2}{x^2}\right) \qquad \text{for~} x\neq 0.
\end{equation}

These bounds are shown in \Cref{fig:HS_comparison} together with other common probability distributions (standardized). Note that \yq{the horseshoe prior has heavy tails that are comparable to the Cauchy distribution,} and it has a pole at $x=0$. These characteristics typically enable the prior to perform well when handling sparsity. We exploit this behavior to develop a horseshoe-based model that can be applied to a class of linear Bayesian inverse problems where edge-preservation is fundamental.
\begin{figure}[!ht]
\centering
\includegraphics[width=0.65\textwidth]{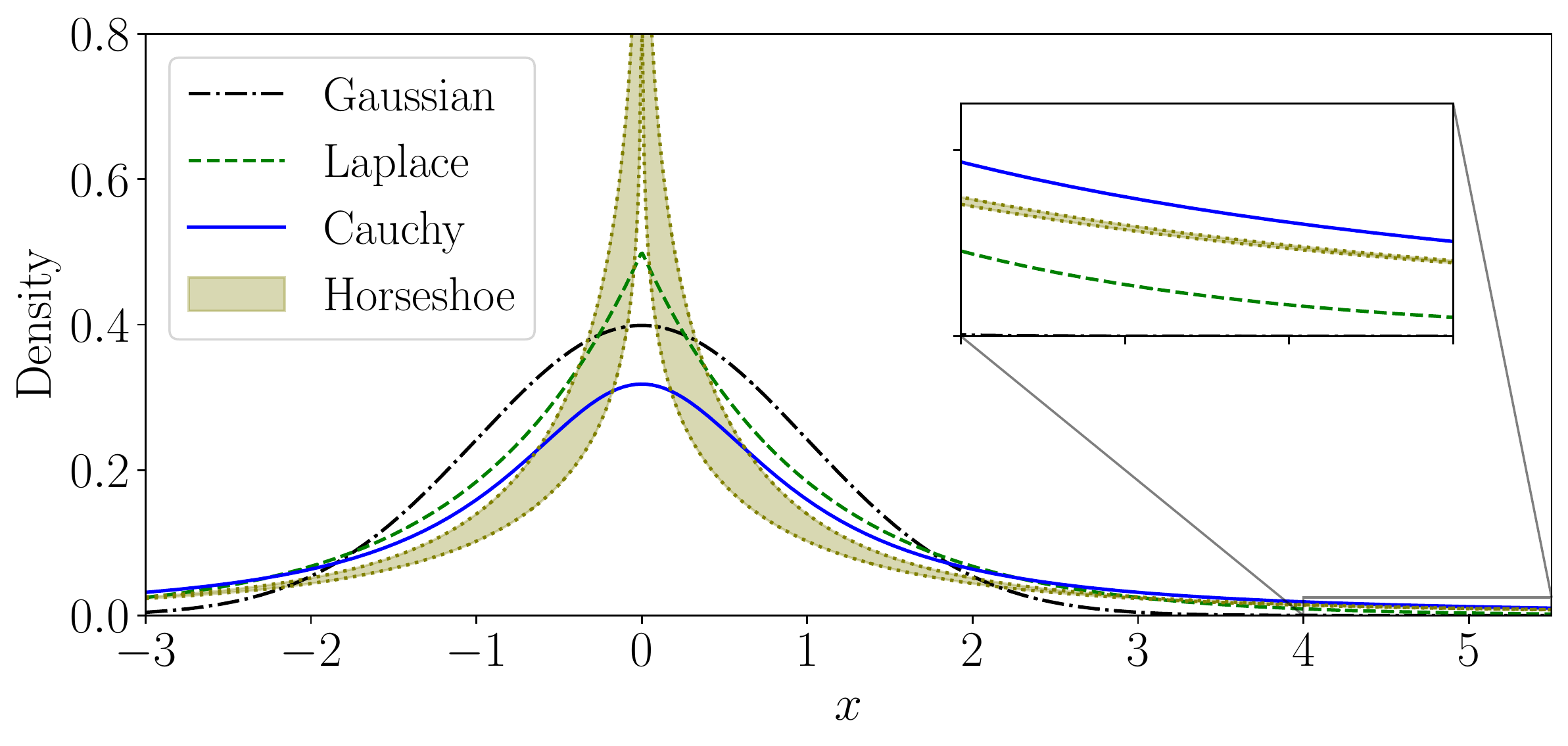}
\caption{Comparison of the horseshoe prior bounds (shaded area) with other common probability densities. The zoom-in highlights the distributions at the tails.}
\label{fig:HS_comparison}
\end{figure}

\section{Horseshoe prior for edge-preservation}\label{sec:HSedge}
In some applications the solution of inverse problems is sparse; in other applications sharp edges are present in the solution meaning that pairwise differences or increments of the solution elements exhibit sparsity. We exploit this fact to define a horseshoe prior on the differences that promote solutions preserving their sharp features. The resulting prior can be interpreted as an anisotropic conditionally Gaussian Markov random field prior.

\subsection{Horseshoe prior and conditionally Gaussian Markov random fields}
Gaussian Markov random fields (GMRF) are typically specified by their conditional independence structure. Therefore, a natural way to describe a GMRF is by its precision matrix. This is because the sparsity structure of this matrix determines the so-called neighborhood system explaining conditional relations in $\ve{x}$ \cite{rue_and_held_2005}. Since we are interested in preserving the edges, we define the precision matrix of the GMRF using increments between elements of the parameter vector. In one-dimensional cases, the increments are computed by application of a finite difference matrix $\mat{D}$, while in two-dimensional cases we consider the horizontal and vertical first-order finite difference matrices $\mat{D}^{(1)},\mat{D}^{(2)}$. These are given by
\begin{equation}\label{eq:diff}
\mat{D}=\begin{bmatrix}
	   1 &   &        &    &   \\
      -1 &  1 &        &    &   \\
         & -1 &      1 &    &   \\
         &    & \ddots & \ddots &   \\
         &    &        & -1 & 1 \\
    \end{bmatrix}_{N\times N}\quad\text{and}\quad  \mat{D}^{(1)} = \mat{I}_N\otimes \mat{D},\quad \mat{D}^{(2)} = {\mat{D}} \otimes \mat{I}_N,
\end{equation}
where $N$ is domain discretization size,
$\mat{I}_N$ is the identity matrix of \yq{size} $N$, and $\otimes$ denotes the Kronecker product. Note that we use, without loss of generality, zero boundary conditions on $\ve{x}$ at the left part of the domain.

Due to the structure defined in \cref{eq:diff} the increments of $\ve{x}$ (i.e., the vectors $\mat{D}\ve{x}$, $\mat{D}^{(1)}\ve{x}$ and $\mat{D}^{(2)}\ve{x}$) are independent and we impose a horseshoe prior on them. This allows us to rewrite the prior in \cref{eq:pr_x} as a conditionally GMRF that incorporates information about the increments. The associated probability density is
\begin{equation}\label{eq:pr_x2}
\pi_{\mathrm{pr}}(\ve{x}\given\tau,\ve{w}) =\frac{(\det\mat{\Lambda}(\tau,\ve{w}))^{\nicefrac{1}{2}}}{(2\pi)^{\nicefrac{d}{2}}} \exp\left(-\frac{1}{2}\ve{x}^\tran \mat{\Lambda}(\tau,\ve{w}) \,\ve{x}\right),
\end{equation}
where $\mat{\Lambda}(\tau,\ve{w})$ is a prior {precision matrix} (see below) that depends on the global standard deviation $\tau\in\mathbb{R}_{>0}$ and local weights $\ve{w}=[w_1,\ldots,w_d]^\tran\in\mathbb{R}^d_{>0}$. These hyperparameters are defined as in the standard horseshoe prior and they are half-Cauchy-distributed following \cref{eq:hyperpr}. In this case, the weights in $\ve{w}$ are analogous to local standard deviations in $\ve{\sigma}$, but they will correspond to the increments and not to the elements of the parameter vector, hence the different notation.

For the conditionally Gaussian prior in \cref{eq:pr_x2}, the precision matrix is defined such that it takes into account the increments via \cref{eq:diff} (see, e.g., \cite[p.68]{bardsley_2019} for the details).
Therefore, we can generally write the precision matrix as
\begin{equation}\label{eq:HS_prec}
\mat{\Lambda}(\tau,\ve{w}) = \mat{L}^\tran\mat{W}(\tau,\ve{w})\mat{L},
\end{equation}
and for one-dimensional and two-dimensional domains we have
\begin{equation}\label{eq:HS_weig1D}
\mat{L}=\mat{D} , \quad
\yq{\mat{W}(\tau,\ve{w})=
\begin{bmatrix}
  \frac{1}{\tau^2w_1^2} &   & \\
   &   \ddots & \\
   &    & \frac{1}{\tau^2w_d^2}
\end{bmatrix}} 
\end{equation}
and
\begin{equation}\label{eq:HS_weig2D}
\mat{L}=\begin{bmatrix}
\mat{D}^{(1)}\\
\mat{D}^{(2)}
\end{bmatrix} , \quad
\mat{W}(\tau,\ve{w})=\begin{bmatrix}
\mat{W}^{(1)} &\mat{0}_d\\
\mat{0}_d & \mat{W}^{(2)}
\end{bmatrix}, \qquad
\yq{\mat{W}^{(i)}=
\begin{bmatrix}
  \frac{1}{\tau^2(w_1^{(i)})^2} &  &  \\
   &   \ddots & \\
   &   & \frac{1}{\tau^2(w_d^{(i)})^2}
\end{bmatrix} }, 
\quad i = 1,2 ,
\end{equation}
respectively. Here, $\mat{L}$ is called the structure matrix, and $\mat{0}_d\in\mathbb{R}^{d\times d}$ is a matrix containing zeros. Notice that in the two-dimensional case we have to consider local weights
\[
   \ve{w} = \begin{bmatrix} \ve{w}^{(1)} \\ \ve{w}^{(2)} \end{bmatrix}
   \in\mathbb{R}^{2d}_{>0}
\]
for each coordinate direction. These are assumed independent and identically distributed according to the half-Cauchy prior exactly as in the one-dimensional case.

The conditionally Gaussian prior \cref{eq:pr_x2} (based on precision matrices of the form \cref{eq:HS_prec} with \cref{eq:HS_weig1D}--\cref{eq:HS_weig2D}) and the hyperpriors \cref{eq:hyperpr} define a horseshoe prior promoting edge-preservation. Again, this is because we are now regularizing \emph{increments} of the unknown parameter using a heavy-tailed probabilistic model. A main motivation to utilize the horseshoe prior is that the unknown parameter is conditionally Gaussian, which facilitates the analytical and numerical treatment of the resulting posterior distribution. However, the main difficulties in applying this prior are: (i) the dimension of the parameter space is increased since the local weights have the same or \yq{even} larger dimension than the model parameter, and (ii) the hyperparameters are endowed with heavy-tailed distributions that cause challenges when computing the posterior via sampling methods. For the latter point, a \emph{regularized horseshoe prior} is proposed in \cite{piironen_and_vehtari_2017}; this model employs half-Student's t-distributions with larger degrees of freedom on the local parameters to overcome the sampling issues. However, increasing the degrees of freedom makes the tail of the distribution less heavy, which is not desirable when looking for methods that promote sharp features. 

\subsection{Extended horseshoe prior}\label{subsec:extendedHS}
Our objective is to present an equivalent model that extends the hierarchical structure of the horseshoe prior by adding auxiliary parameters. This allows us to write the resulting posterior such that it can be sampled in a tractable manner without loosing the connection to the original half-Cauchy hyperpriors in the standard horseshoe. We point out that this idea is also used in \cite{makalic_and_schmidt_2016} for the standard horseshoe prior applied to logistic regression problems.

We consider a scale mixture decomposition of a Student's t-dis\-tri\-buted random variable (see, e.g., \cite{wand_et_al_2011}). If $A$ and $B$ are random variables such that
\begin{equation}\label{eq:decomp}
    (A^2\given B) \sim \invg\left(\frac{\nu}{2}, \frac{\nu}{B}\right) \quad \text{and} \quad
    B \sim \invg\left(\frac{1}{2}, \frac{1}{c^2}\right),
\end{equation}
where $\invg(\cdot,\cdot)$ denotes the inverse gamma distribution depending on shape and scale parameters, then
\begin{equation}
    A\sim\stud^+(\nu,0, c),
\end{equation}
where $\stud^+(\cdot,\cdot,\cdot)$ is the half-Student's t-distribution depending on degrees of freedom, location and scale parameters. These distributions have densities defined as
\begin{subequations}
\begin{align}
\invg(x; \alpha,\beta) &= \frac{\beta^\alpha}{\Gamma(\alpha)}x^{-\alpha-1}\exp\left(-\frac{\beta}{x}\right),\\
\stud^+(x; \nu,0, c) &= \left\{ \begin{array}{ll}
  \dfrac{2\Gamma\left({\frac{\nu +1}{2}}\right)}{\Gamma\left({\frac{\nu }{2}}\right){\sqrt{\pi\nu}}\,c\,}\left(1+{\frac{1}{\nu}}\frac{x^2}{c^2}\right)^{-{\frac{\nu+1}{2}}},
  & \text{if $x\geq0$} \\[4mm] 0, & \text{otherwise.} \end{array} \right.
\end{align}
\end{subequations}

The decomposition \cref{eq:decomp} can be derived by marginalization of the associated hierarchical prior:
\begin{subequations}\label{eq:parentdecomp}
\begin{align}
\pi(a^2) &= \int_0^{\infty} \pi(a^2\given b)\pi(b)\,\dd b
\,\propto\, (a^2)^{-\frac{\nu}{2}-1} \int_0^{\infty} {b^{-\frac{\nu+1}{2}-1}} \exp\left(-\frac{1}{b}\left(\frac{\nu}{a^2}+\frac{1}{c^2}\right)\right) \dd b
\label{eq:decomp1}\\
&\propto (a^2)^{-\frac{\nu}{2}-1}{\left(\frac{\nu}{a^2}+\frac{1}{c^2}\right)^{-\frac{\nu+1}{2}}} \propto (a^2)^{-\frac{1}{2}} \left(1+\frac{1}{\nu}\frac{a^2}{c^2}\right)^{-\frac{\nu+1}{2}},\label{eq:decomp2}
\end{align}
\end{subequations}
where we use the fact that the integrand in \cref{eq:decomp1} is an unnormalized inverse gamma density with shape and scale parameters ${(\nu+1)}/{2}$ and ${\nu}/{a^2}+{1}/{c^2}$, respectively. This allows analytical computation of the marginalization procedure. Finally, to obtain the density for $A$, we can use a probabilistic transformation (as performed for \cref{eq:kappa}). This results in the half-Student's t-distribution defined in \cref{eq:decomp} with degrees of freedom $\nu$ and scale parameter $c$. We consider the standard horseshoe prior and therefore we use degrees of freedom $\nu=1$ such that half-Student's t-distribution reduces to a half-Cauchy distribution. In addition, we note that the regularized horseshoe prior \cite{piironen_and_vehtari_2017}, which typically uses $\nu=3$, can also be included in the formulation.

The scale mixture representation \cref{eq:decomp} is utilized to write the standard horseshoe prior as an extended hierarchical prior. The motivation behind this choice is that the additional auxiliary variables allow the derivation of full conditional distributions in closed form by exploiting conjugacy. Thereafter, these can be sampled within a Gibbs sampler scheme \cite[Ch.10]{robert_and_casella_2004}. The resulting \emph{extended horseshoe prior} is defined as:
\begin{equation}\label{eq:hHS_prior}
\pi_{\mathrm{pr}}(\ve{x},\tau^2,\ve{w}^2,\gamma,\ve{\xi}) = \pi_{\mathrm{pr}}(\ve{x}\given \tau^2,\ve{w}^2)\pi_{\mathrm{hpr}}(\tau^2\given \gamma)\pi_{\mathrm{hpr}}(\gamma)\pi_{\mathrm{hpr}}(\ve{w}^2\given \ve{\xi})\pi_{\mathrm{hpr}}(\ve{\xi}),
\end{equation}
where
\begin{subequations}\label{eq:hHS_priorcomp}
\begin{align}
  \pi_{\mathrm{pr}}(\ve{x}\given \tau^2,\ve{w}^2) &= \mathcal{N}\left(\ve{0},
    \mat{\Lambda}^{-1}(\tau,\ve{w})\right), \\
  \pi_{\mathrm{hpr}}(\tau^2\given \gamma) &= \invg\left(\frac{\nu}{2},
    \frac{\nu}{\gamma}\right), & 
    \pi_{\mathrm{hpr}}(\gamma) &= \invg\left(\frac{1}{2}, \frac{1}{\tau_0^2}\right),\\
  \pi_{\mathrm{hpr}}({w}_i^2\given {\xi}_i) &= \invg\left(\frac{\nu}{2},
    \frac{\nu}{\xi_i}\right), & 
    \pi_{\mathrm{hpr}}({\xi}_i) &= \invg\left(\frac{1}{2}, 1\right). \label{eq:hHS_priorb}
\end{align}
\end{subequations}


\section{Proposed approach}\label{sec:method}
The extended hierarchical horseshoe prior \cref{eq:hHS_prior} is used to define the posterior in \cref{eq:Bayes}, thereby defining a hierarchical formulation of the Bayesian inverse problem. In addition to the hyperparameters arising from the horseshoe prior, we also model the noise variance $\sigma_{\mathrm{obs}}^2$ in the likelihood function as a random variable. Following the hyperprior distributions imposed on the prior parameters, we define an inverse gamma hyperprior for $\sigma_{\mathrm{obs}}^2$ with shape and scale parameters $\alpha_{\mathrm{obs}}=1$ and $\beta_{\mathrm{obs}}=10^4$, respectively. This choice makes the hyperprior relatively uninformative \cite{higdon_2007}.

As a result, the Bayesian inverse problem of estimating the posterior \cref{eq:Bayes} is written as the hierarchical Bayesian inverse problem of determining the new posterior density:
\begin{multline}\label{eq:Bayes_hrc}
\pi_{\mathrm{pos}}(\ve{x},\sigma_{\mathrm{obs}}^2,\tau^2,\ve{w}^2,\gamma,\ve{\xi}) \propto \\
\pi_{\mathrm{data}}(\ve{y}\given\ve{x},\sigma_{\mathrm{obs}}^2) \pi_{\mathrm{pr}}(\ve{x}\given\tau^2,\ve{w}^2) \pi_{\mathrm{hpr}}(\tau^2\given\gamma) \pi_{\mathrm{hpr}}(\ve{w}^2\given\ve{\xi}) \pi_{\mathrm{hpr}}(\sigma_{\mathrm{obs}}^2) \pi_{\mathrm{hpr}}(\gamma) \pi_{\mathrm{hpr}}(\ve{\xi});
\end{multline}
whose dependencies are shown in \Cref{fig:HS_hierarchies}.
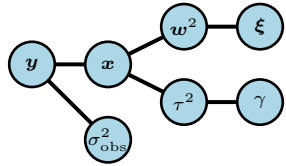
\begin{figure}[!ht]
\centering
\begin{tikzpicture}
\Vertex[x=4,label=$\ve{y}$]{E} \Vertex[x=5,label=$\ve{x}$]{F} \Vertex[x=6,y=0.5,label=$\ve{w}^2$]{G}\Vertex[x=6,y=-0.5,label=$\tau^2$]{H}\Vertex[x=7,y=0.5,label=$\ve{\xi}$]{I}\Vertex[x=7,y=-0.5,label=${\gamma}$]{J}\Vertex[x=5,y=-1.0,label=$\sigma_{\mathrm{obs}}^2$]{K}
\Edge[color=black](E)(F)\Edge[color=black](F)(G)\Edge[color=black](F)(H)\Edge[color=black](G)(I)\Edge[color=black](H)(J)\Edge[color=black](E)(K)
\end{tikzpicture}
\caption{Structure of the hierarchical Bayesian inference model \eqref{eq:Bayes_hrc} based on the extended horseshoe prior.}
\label{fig:HS_hierarchies}
\end{figure}

Performing statistical inference with the posterior \cref{eq:Bayes_hrc} generally requires application of sampling methods based on Markov chain Monte Carlo (MCMC) methods. The goal in MCMC is to compute samples or realizations of a Markov chain that is stationary with respect to the posterior distribution \cite{owen_2019}. A common difficulty when solving hierarchical Bayesian inverse problems is that MCMC algorithms have to handle drastic changes in the shape of the posterior. This is because a small modification at the bottom of the hierarchy induces large changes in the joint posterior (see, e.g., \cite{betancourt_and_girolami_2015} for a detailed discussion). To alleviate this problem one typically relies on (i) re-parametrization of the hierarchical structure in order to break potential correlations between parameters (cf. \Cref{fig:HS_hierarchies}), or (ii) application of specialized MCMC algorithms that capture the geometry of the joint posterior. For instance, a Riemannian Hamiltonian Monte Carlo method that uses local curvature information is advocated in \cite{betancourt_and_girolami_2015}; and for problems involving the horseshoe prior, a Gibbs sampler is proposed in \cite{makalic_and_schmidt_2016} and two further MCMC methods are developed in \cite{johndrow_et_al_2020}. Recently, another Gibbs sampler is discussed in \cite{nishimura_and_suchard_2022} for Bayesian inverse problems under the regularized horseshoe prior.

Our idea is to exploit the structure presented in \yq{section \ref{subsec:extendedHS}} to derive full conditional distributions for each uncertain parameter. The advantage of this approach is that it allows a direct application of the Gibbs sampler \cite{geman_and_geman_1984}, since the conditional densities for each parameter can be derived in closed form. Therefore, we avoid sampling in the high-dimensional joint space of the posterior. In particular, the full conditional densities associated with \cref{eq:Bayes_hrc} are found to be:
\begin{subequations}\label{eq:conds}
\begin{align}
\pi_1\left(\ve{x}\given \sigma_{\mathrm{obs}}^2,\tau^2,\ve{w}^2\right) &\propto \pi_{\mathrm{data}}(\ve{y}\given\ve{x},\sigma_{\mathrm{obs}}^2) \pi_{\mathrm{pr}}(\ve{x}\given\tau^2,\ve{w}^2),\label{eq:conds1}\\
\pi_2(\sigma_{\mathrm{obs}}^2\given \ve{x}) &\propto \pi_{\mathrm{data}}(\ve{y}\given\ve{x},\sigma_{\mathrm{obs}}^2)\pi_{\mathrm{hpr}}(\sigma_{\mathrm{obs}}^2),\label{eq:conds2}\\
\pi_3(\tau^2\given \ve{x}, \ve{w}^2, \gamma) &\propto \pi_{\mathrm{pr}}(\ve{x}\given\tau^2,\ve{w}^2)\pi_{\mathrm{hpr}}({\tau}^2\given{\gamma}),\label{eq:conds3}\\
\pi_4(\ve{w}^2\given \ve{x}, \tau^2, \ve{\xi}) &\propto \pi_{\mathrm{pr}}(\ve{x}\given\tau^2,\ve{w}^2)\pi_{\mathrm{hpr}}(\ve{w}^2\given\ve{\xi}),\label{eq:conds4}\\
\pi_5(\gamma\given \tau^2) &\propto \pi_{\mathrm{hpr}}({\tau}^2\given{\gamma})\pi_{\mathrm{hpr}}(\gamma),\label{eq:conds5}\\
\pi_6(\ve{\xi}\given \ve{w}^2) &\propto \pi_{\mathrm{hpr}}(\ve{w}^2\given\ve{\xi})\pi_{\mathrm{hpr}}(\ve{\xi}).\label{eq:conds6}
\end{align}
\end{subequations}

In the remainder of this section, we determine the full conditional densities for each uncertain parameter and discuss sampling techniques to obtain draws from them. Due to the extended horseshoe model, we anticipate that most of the densities in \cref{eq:conds} can be sampled directly and a classic Gibbs sampler algorithm can be used to characterize the posterior.

\subsection{Sampling of $\pi_1$}\label{subsec:x_cgls}
The conditional for the uncertain parameter $\ve{x}$ in \cref{eq:conds1} defines a Bayesian inverse problem with Gaussian likelihood and prior. After inserting the corresponding probabilistic models we obtain:
\begin{equation}\label{eq:cond1}
\pi_1\left(\ve{x}\given \sigma_{\mathrm{obs}}^2,\tau^2,\ve{w}^2\right) \propto \exp\left(-\frac{1}{2} \left(\frac{1}{\sigma_{\mathrm{obs}}^2}\norm{\ve{y}-\mat{A}\ve{x}}_2^2 + \bigl\| \mat{\Lambda}^{\nicefrac{1}{2}}(\tau,\ve{w})\,\ve{x} \bigr\|_2^2\right) \right).
\end{equation}
This conditional density is also Gaussian with precision matrix and mean vector given by (see, e.g., \cite[p.78]{kaipio_and_somersalo_2005})
\begin{equation}\label{eq:postparams}
\widetilde{\mat{\Lambda}}(\tau,\ve{w}) =  \frac{1}{\sigma_{\mathrm{obs}}^2}\mat{A}^\tran\mat{A} + \mat{\Lambda}(\tau,\ve{w}), \qquad \widetilde{\ve{\mu}}(\tau,\ve{w}) = \widetilde{\mat{\Lambda}}^{-1}(\tau,\ve{w})\left(\frac{1}{\sigma_{\mathrm{obs}}^2}\mat{A}^\tran\ve{y}\right).
\end{equation}
where $\mat{\Lambda}(\tau,\ve{w})$ is given in \eqref{eq:HS_prec}.

Given the hyperparameters $\sigma_{\mathrm{obs}}$, $\tau$ and $\ve{w}$, the most common sampling algorithm for a Gaussian distribution is based on the Cholesky factorization. In this case, a sample from $\pi_1$ is obtained as $\ve{x}^\star= \widetilde{\ve{\mu}} + \widetilde{\mat{\Lambda}}^{-\nicefrac{1}{2}}\ve{u}$, where $\ve{u}\sim\mathcal{N}(\ve{0},\mat{I}_d)$ is a standard Gaussian random vector, and $\widetilde{\mat{\Lambda}}^{\nicefrac{1}{2}}$ is a lower triangular matrix with real and positive diagonal entries (Cholesky factor). Note that we drop the dependence on $\sigma_{\mathrm{obs}}$, $\tau$ and $\ve{w}$ for the sake of notation. This simple strategy is computationally prohibitive due to the need for computing a factorization of the matrix $\widetilde{\mat{\Lambda}}(\tau,\ve{w})$. Assuming that matrix-vector multiplications with $\widetilde{\mat{\Lambda}}$ can be done efficiently, we rely on Krylov subspace methods to sample from the Gaussian conditional in \cref{eq:cond1}. We refer to \cite{vono_et_al_2022} for a review of methods for sampling high-dimensional Gaussian distributions. In particular, the task of sampling a Gaussian random vector can be written as a least squares problem, and thus, we draw a sample $\ve{x}^\star$ from $\pi_1$ by solving:
\begin{equation}\label{eq:CGLS}
  \ve{x}^\star = \argmin_{\ve{x}\in \mathbb{R}^{d}}
    \norm{ \mat{M} \ve{x} - \ve{z} }_2^2
    \quad \hbox{with} \quad \mat{M} =
    \begin{bmatrix}
      ({\nicefrac{1}{\sigma_{\mathrm{obs}}}})\mat{A}\\
      \mat{\Lambda}^{\nicefrac{1}{2}}
    \end{bmatrix} , \quad \ve{z} =
    \begin{bmatrix}
      ({\nicefrac{1}{\sigma_{\mathrm{obs}}}})\ve{y} \\
      \ve{0}_{d}
    \end{bmatrix} + \widetilde{\ve{u}} ,
\end{equation}
where $\widetilde{\ve{u}}\sim\mathcal{N}(\ve{0},\mat{I}_{m+d})$. Recall that the solution of \cref{eq:CGLS} is required at every Gibbs iteration given new values of the hyperparameters. We use the CGLS method, which requires one pair of forward and backward model computations (i.e., multiplications with $\mat{A}$ and $\mat{A}^\tran$) per iteration. A tolerance $\varepsilon_{\mathrm{cgls}}$ on the relative residual norm of the normal equations or a maximum number of iterations $n_{\max}$ can be used to control the quality of the computed samples.

One way to accelerate the performance of the CGLS method is to apply a standard-form transformation to \eqref{eq:CGLS} \cite[Sec.~8.4]{hansen_2010}; this is also referred to as \emph{priorconditioning} \cite{calvetti_et_al_2018}. The idea is to compute the Cholesky factorization of the prior precision matrix $\mat{\Lambda}=\mat{C}^\tran\mat{C}$ and introduce the change of variables $\widetilde{\ve{x}}=\mat{C}\ve{x}$, such that we can transform \cref{eq:CGLS} into
\begin{equation}\label{eq:CGLS_st}
  \tilde{\ve{x}}^\star = \argmin_{\tilde{\ve{x}}\in \mathbb{R}^{d}}
    \norm{ \mat{M} \widetilde{\ve{x}} - \ve{z} }_2^2
    \quad \hbox{with} \quad \mat{M} =
    \begin{bmatrix}
      ({\nicefrac{1}{\sigma_{\mathrm{obs}}}})\mat{A}\mat{C}^{-1}\\
      \mat{I}_d
    \end{bmatrix},
\end{equation}
where $\ve{z}$ is defined as in \cref{eq:CGLS}. Note that the change of variables corresponds to ``whitening'' the prior, that is, $\widetilde{\ve{x}}\sim \mathcal{N}(\ve{0},\mat{I}_d)$. In the standard-form least squares problem \eqref{eq:CGLS_st}, we are now working with the Krylov subspace $\mathrm{span}\{\mat{\Lambda}^{-1}\mat{A}^\tran\ve{y}, (\mat{\Lambda}^{-1}\mat{A}^\tran\mat{A}) \allowbreak\mat{\Lambda}^{-1}\mat{A}^\tran\ve{y}, (\mat{\Lambda}^{-1}\mat{A}^\tran\mat{A})^2\mat{\Lambda}^{-1}\mat{A}^\tran\ve{y}, \ldots\}$, which is in most cases a better subspace than $\mathrm{span}\{\mat{A}^\tran\ve{y}, (\mat{A}^\tran\mat{A})\mat{A}^\tran\ve{y}, \allowbreak(\mat{A}^\tran\mat{A})^2\mat{A}^\tran\ve{y}, \ldots\}$ of the general-form least-squares problem \cref{eq:CGLS} (see, e.g., \cite[Sec.~8.4]{hansen_2010} for the details). Hence, the application of the standard-form transformation tends to reduce the number of iterations required for convergence and it can be incorporated in a preconditioned version of the CGLS algorithm (\cite[Sec.~7.4]{bjorck_1996}). We denote this variant as pCGLS.

The Cholesky factor of the prior precision matrix, which is instrumental for the standard-form transformation, can be obtained as follows. In the one-dimensional case, from the definition of the prior precision in \eqref{eq:HS_weig1D} we have that
  \begin{equation}
    \mat{\Lambda} = \mat{L}^\tran\mat{W}\mat{L} = \mat{C}^\tran\mat{C} \qquad \text{with}\quad  \mat{L}=\mat{D},
  \end{equation}
where $\mat{W}$ is diagonal and $\mat{C}$ is either upper or lower triangular. Hence, by simple observation, we can write $\mat{C} = \mat{W}^{\nicefrac{1}{2}}\mat{D}$, which is indeed triangular and constitutes an inexpensive representation of the Cholesky factor.

In the two-dimensional case, the Cholesky factor is no longer immediately available but still can be computed in an economical way. From \cref{eq:HS_weig2D} we have
\begin{equation}
\mat{\Lambda}= \mat{L}^\tran{\mat{W}}\mat{L}\qquad \text{with}\qquad
{\mat{W}}=
\begin{bmatrix}
\mat{W}^{(1)} &\ve{0}_d\\
\ve{0}_d & \mat{W}^{(2)}
\end{bmatrix}\quad \text{and}\quad
\mat{L}=\begin{bmatrix}
\mat{D}^{(1)}\\
\mat{D}^{(2)}
\end{bmatrix},
\end{equation}
where the difference matrices $\mat{D}^{(1)}$ and $\mat{D}^{(2)}$ are sparse and so is $\mat{L}$. Now, we can compute the thin QR factorization of ${\mat{W}}^{\nicefrac{1}{2}}\mat{L}$,
\begin{equation}\label{eq:QR}
{\mat{W}}^{\nicefrac{1}{2}}\mat{L}=\begin{bmatrix}
(\mat{W}^{(1)})^{\nicefrac{1}{2}}\mat{D}^{(1)}\\
(\mat{W}^{(2)})^{\nicefrac{1}{2}}\mat{D}^{(2)}
\end{bmatrix} = \mat{Q}\mat{R};
\end{equation}
notice that $\mat{Q}$ is not required and $\mat{L}^\tran{\mat{W}}\mat{L}=\mat{R}^\tran\mat{R}$. Therefore, $\mat{C} =\mat{R}$ is the desired Cholesky factor.

\subsection{Sampling of $\pi_2$ to $\pi_6$}\label{sec:samp_hyper}
The conditional densities for each hyperparameter are all obtained in closed form. This is due to conjugate relations that arise from the extended horseshoe prior formulation. 

The conditional density for the \emph{noise variance} in \cref{eq:conds2} is
\begin{subequations}\label{eq:cond_sigmaobs2}
\begin{align}
\pi_2(\sigma_{\mathrm{obs}}^2\given \ve{x}) &\propto \pi_{\mathrm{data}}(\ve{y}\given\ve{x},\sigma_{\mathrm{obs}}^2)\pi_{\mathrm{hpr}}(\sigma_{\mathrm{obs}}^2)\\ &\propto (\sigma_{\rm obs}^2)^{-\nicefrac{m}{2}} \exp\left(-\frac{1}{2\sigma_{\rm obs}^2}\norm{\ve{y}-\mat{A}\ve{x}}_2^2\right)\left[(\sigma_{\mathrm{obs}}^2)^{-\alpha_{\mathrm{obs}}-1}\exp\left(-\frac{1}{\beta_{\mathrm{obs}}\sigma_{\mathrm{obs}}^2}\right)\right]\\
&\propto (\sigma_{\rm obs}^2)^{-\nicefrac{m}{2}-\alpha_{\mathrm{obs}}-1}\exp\left(-\frac{1}{\sigma_{\rm obs}^2}\left[\frac{1}{2}\norm{\ve{y}-\mat{A}\ve{x}}_2^2+\frac{1}{\beta_{\mathrm{obs}}}\right]\right) \\
&\propto \invg\left(\frac{m}{2}+\alpha_{\mathrm{obs}},\, \frac{1}{2}\norm{\ve{y}-\mat{A}\ve{x}}_2^2 + \frac{1}{\beta_{\mathrm{obs}}}\right).\label{eq:cond_sigmaobs2b}
\end{align}
\end{subequations}

The conditional density for the squared \emph{global shrinkage parameter} in \cref{eq:conds3} is
\begin{subequations}
\begin{align}
\pi_3(\tau^2\given \ve{x}, \ve{w}^2, \gamma) &\propto \pi_{\mathrm{pr}}(\ve{x}\given\tau^2,\ve{w}^2)\pi_{\mathrm{hpr}}({\tau}^2\given{\gamma})\\ &\propto (\tau^2\,\ve{w}^2)^{-\nicefrac{k}{2}} \exp\left(-\frac{1}{2} \sum_{i=1}^{k}\frac{[\mat{L}\ve{x}]_i^2}{\tau^2\,w_i^2} \,\right) \left[(\tau^2)^{-\nicefrac{\nu}{2}-1}\exp\left(-\frac{\nu}{\gamma\tau^2}\right)\right]\\
&\propto (\tau^2)^{-\nicefrac{k}{2}-\nicefrac{\nu}{2}-1} \exp\left(-\frac{1}{\tau^2}\left[\frac{1}{2}\sum_{i=1}^{k}\frac{[\mat{L}\ve{x}]_i^2}{w_i^2}+\frac{\nu}{\gamma}\right]\right)\\
&\propto\invg\left(\frac{k+\nu}{2},\, \frac{1}{2}\sum_{i=1}^{k}\frac{[\mat{L}\ve{x}]_i^2}{w_i^2}+\frac{\nu}{\gamma}\right),\label{eq:cond_tau2}
\end{align}
\end{subequations}
where $k=\{d, 2d\}$ in one- and two-dimensional problems, respectively.

Moreover, since the \emph{local shrinkage parameters} are assumed independent, one can derive the conditional density for their squared version and at each component \cref{eq:conds4}, as follows
\begin{subequations}
\begin{align}
\pi_4(w_i^2\given \ve{x}, \tau^2,\xi_i) &\propto \pi_{\mathrm{pr}}(\ve{x}\given\tau^2,w_i^2)\pi_{\mathrm{hpr}}(w_i^2\given{\xi_i})\\ &\propto (\tau^2\,w_i^2)^{-\nicefrac{1}{2}} \exp\left(-\frac{1}{2} \frac{[\mat{L}\ve{x}]_i^2}{\tau^2\,w_i^2} \,\right) \left[(w_i^2)^{-\nicefrac{\nu}{2}-1}\exp\left(-\frac{\nu}{\xi_i\,w_i^2}\right)\right]\\
&\propto (w_i^2)^{-\nicefrac{1}{2}-\nicefrac{\nu}{2}-1} \exp\left(-\frac{1}{w_i^2}\left[\frac{[\mat{L}\ve{x}]_i^2}{2\tau^2}+\frac{\nu}{\xi_i}\right]\right)\\
&\propto\invg\left(\frac{\nu+1}{2},\, \frac{[\mat{L}\ve{x}]_i^2}{2\tau^2}+\frac{\nu}{\xi_i}\right).\label{eq:cond_w2}
\end{align}
\end{subequations}

The conditionals for the \emph{auxiliary parameters} $\gamma$ and $\ve{\xi}$ in \cref{eq:conds5} and \cref{eq:conds6} respectively, are derived in a similar manner. These are inverse gamma distributions defined as:
\begin{subequations}\label{eq:conds_g_x}
\begin{align}
\pi_5\left(\gamma\given \tau^2\right) &\propto\invg\left(\frac{\nu+1}{2},\,  \frac{1}{\tau_0^2}+\frac{\nu}{\tau^2}\right),\label{eq:cond_gamma}\\
\pi_6\left(\xi_i\given w_i^2\right) &\propto\invg\left(\frac{\nu+1}{2},\, 1+\frac{\nu}{w_i^2}\right).\label{eq:cond_xi}
\end{align}
\end{subequations}

\subsection{The computational procedure}\label{subsec:full}
Based on the sampling approaches for the full conditional densities discussed above, we define a Gibbs sampler generating $n_s$ states of a Markov chain with stationary distribution \cref{eq:Bayes_hrc}, $\{\ve{x}^{(j)}, ({\sigma}_{\mathrm{obs}}^2)^{(j)}, (\tau^2)^{(j)},\allowbreak (\ve{w}^2)^{(j)}, \allowbreak\gamma^{(j)}, \ve{\xi}^{(j)}\}_{j=1}^{n_s}$.

The Gibbs sampler draws a single sample from each conditional density at each iteration. This uses the fact that, under mild conditions, the set of full conditional distributions determine the joint distribution \cite{besag_1974}. The Markov chain approaches its equilibrium condition as the number of iterations increases. This means that after convergence all samples from the chain will be distributed according to the target posterior distribution. Convergence conditions for the Gibbs sampler are defined in \cite{roberts_and_smith_1994, tierney_1994}. From a practical viewpoint, convergence is guaranteed approximately and explored from a statistical perspective by analyzing the observed output of the chain and exploring ergodic results. For this, we require application of burn-in and lagging steps: (i) the $n_s$ successive values of the chain are only selected after discarding $n_b$ samples during the warm up phase of the algorithm (burn-in period), and (ii) the sample chain values are only stored every $n_t$ iterations since for large enough $n_t$ the samples are virtually independent (lagging steps). As a result, to obtain $n_s$ quasi-independent samples from the posterior, we require $n=n_b+n_sn_t$ Markov chain steps.

Different scanning strategies exist for the Gibbs sampler (see, e.g., \cite{gamerman_and_lopes_2006, owen_2019}). We follow a deterministic or systematic scan in which all iterations consist of sampling the conditional densities of each component in the same order. This version of the Gibbs sampler applied to the posterior  \cref{eq:Bayes_hrc} is summarized in \Cref{algo:gibbs}.
\begin{algorithm}[!ht]
\setcounter{AlgoLine}{0}
\DontPrintSemicolon
\KwIn{conditional densities \cref{eq:conds}, number of samples $n_s$, thinning $n_t$, burn-in $n_b$, \yq{and maximum number of CGLS iterations $n_{\max}$ (or tolerance $\varepsilon_{\mathrm{cgls}}$)}.}

Initial states $\ve{x}^\star, {\sigma}_{\mathrm{obs}}^\star, \tau^\star, \ve{w}^\star, \gamma^\star, \ve{\xi}^\star$, and initial Cholesky factor $\mat{C}(\tau^\star, \ve{w}^\star)$.

\yq{$n\leftarrow n_b+ n_{s}n_t$}, and $j\leftarrow 1$\;

\For{$k = 1,\ldots, n$}{
	\texttt{// Sample target parameter }\;

	$\ve{x}^\star \sim \pi_1(\cdot\given{\sigma}_{\mathrm{obs}}^\star, \tau^\star,\ve{w}^\star)$: solve least-squares problem using preconditioned CGLS with $\mat{C}(\tau^\star, \ve{w}^\star)$\;

	\texttt{// Sample variance hyperparameters}\;

	$\sigma_{\mathrm{obs}}^\star \sim \pi_2\left(\cdot \given \ve{x}^\star\right)$: in closed form \cref{eq:cond_sigmaobs2}

	$\tau^\star \sim \pi_3\left(\cdot \given \ve{x}^\star, \ve{w}^\star,\gamma^\star\right)$: in closed form \cref{eq:cond_tau2}

	$\ve{w}^\star \sim \pi_4\left(\cdot \given \ve{x}^\star,{\tau}^\star, \ve{\xi}^\star\right)$: in closed form \cref{eq:cond_w2}

	\texttt{// Compute Cholesky factor}\;

	$\mat{C}(\tau^\star, \ve{w}^\star)$: based on the approach discussed in \cref{subsec:x_cgls}

	\texttt{// Sample auxiliary hyperparameters}\;

	$\gamma^\star\sim \pi_5\left(\cdot \given \tau^\star\right)$: in closed form \cref{eq:cond_gamma}

	$\ve{\xi}^\star \sim \pi_6\left(\cdot \given \ve{w}^\star\right)$: in closed form \cref{eq:cond_xi}

	\texttt{// Save samples }\;

	\If{$(k > n_b)$}{
	\If{$(k~\mathrm{mod}~n_t)=0$}{
	$\ve{x}^{(j)} \leftarrow \ve{x}^\star,~({\sigma}_{\mathrm{obs}}^2)^{(j)} \leftarrow {\sigma}_{\mathrm{obs}}^\star,~(\tau^2)^{(j)} \leftarrow {\tau}^\star,~(\ve{w}^2)^{(j)} \leftarrow \ve{w}^\star,~{\gamma}^{(j)} \leftarrow \gamma^\star,~\ve{\xi}^{(j)} \leftarrow \ve{\xi}^\star$
	
	$j\leftarrow j+1$\;
	}
	}
}
\Return{$\{\ve{x}^{(j)}, ({\sigma}_{\mathrm{obs}}^2)^{(j)}, (\tau^2)^{(j)}, (\ve{w}^2)^{(j)}, \gamma^{(j)}, \ve{\xi}^{(j)}\}_{j=1}^{n_s}$}
\caption{{\sc Gibbs sampling of the posterior \cref{eq:Bayes_hrc}}}
\label{algo:gibbs}
\end{algorithm}

The efficiency of \Cref{algo:gibbs} can be measured in terms of the autocorrelation of the posterior samples. The chain autocorrelation allows the definition of the \emph{effective sample size} \cite{owen_2019}
\begin{equation}
n_{\text{eff}} := \frac{n_s}{1+2\sum_{j=1}^{n_s}\frac{\rho^{(j)}}{\rho^{(0)}}} \approx \ceil*{\frac{n_s}{\tau_{\mathrm{int}}}},
\end{equation}
where $\rho^{(j)}$ denotes the autocorrelation at the $j$th lag
and $\tau_{\mathrm{int}}$ is the \emph{integrated autocorrelation time} (IACT). Essentially, the effective sample size is used to compare the variance estimated via correlated MCMC samples and the ideal case of a variance computed from independent draws. Thus, the aim is to obtain a value of $n_{\text{eff}}$ as close as possible to $n_s$.

The computational cost of the Gibbs sampler at each iteration can be given in terms of the number of model calls (operations with $\mat{A}$ and $\mat{A}^\tran$). Hence, the cost corresponding to the simulation of each conditional density are: maximum $2{n}_{\max}$ calls are needed for $\pi_1$ and $1$ evaluation is required to sample the density $\pi_2$. Hence, the total cost of a single iteration of the Gibbs sampler is $2n_{\max} + 1$ model calls maximum.

\section{Numerical experiments}\label{sec:numexp}
In the following, we illustrate the use of the horseshoe prior for edge-preserving inversion and the Gibbs sampler in Algorithm~\ref{algo:gibbs} designed for the computations. We consider linear inverse problems arising in imaging science. The first example consists of a one-dimensional deconvolution problem that allows us to test multiple parameter settings in our computational framework. The second example is a two-dimensional computed tomography problem that highlights the potential of the horseshoe prior in more realistic applications.

In our test problems, the point estimates $\bar{\ve{x}}$ (posterior mean or median) of the target parameter are evaluated using the relative reconstruction error defined as
\begin{equation}
	\text{relerr}:=\norm{\bar{\ve{x}}-\ve{x}^{\mathrm{true}}}_2/ \norm{\ve{x}^{\mathrm{true}}}_2,
\end{equation}
where $\ve{x}^{\mathrm{true}}$ denotes the underlying true solution.
To further assess the effectiveness of our approach, we compare our solutions with those from the method proposed in \cite{uribe_et_al_2021} which uses a Laplace Markov random field prior to achieve edge-preservation. This method, however, relies on a Laplace approximation of the posterior, while \Cref{algo:gibbs} does not introduce any approximation.

\subsection{One-dimensional deconvolution}
We consider the inverse problem of identifying an unknown piecewise constant signal $x:[0,1]\to\mathbb{R}$ from noisy convolved data. The mathematical model for convolution can be written as a Fredholm integral equation of the first kind:
\begin{equation}\label{eq:1D_deconv}
y(t) = \int_{0}^{1} \mathcal{A}(t-u)x(u)\,\dd u \quad\text{with}\quad \mathcal{A}(t) = \exp\left(-\frac{1}{2s^2}(t-u)^2\right), \qquad 0\leq t\leq 1,
\end{equation}
where $y(t)$ denotes the convolved signal and we employ a Gaussian convolution kernel $\mathcal{A}(t)$ with fixed parameter $s=0.016$.

In practice, a finite-dimensional representation of \cref{eq:1D_deconv} is employed. After discretizing the signal domain into $d=128$ intervals, the convolution model can be expressed as a system of linear algebraic equations $\ve{y}=\mat{A}\ve{x}$. We consider two sets of synthetic \yq{observed data $\ve{y}$} with $m=d$ equally-spaced elements. The first set is generated with noise standard deviation $\sigma_{\mathrm{obs}}^{\rm true}=7.867\times 10^{-3}$ and the second one with $\sigma_{\mathrm{obs}}^{\rm true}=1.967\times 10^{-2}$ (corresponding to $2\%$ and $5\%$ errors). \Cref{fig:ex_deconv_data} shows the true signal together with the two data sets.
\begin{figure}[!ht]
\centering
\includegraphics[width=0.99\textwidth]{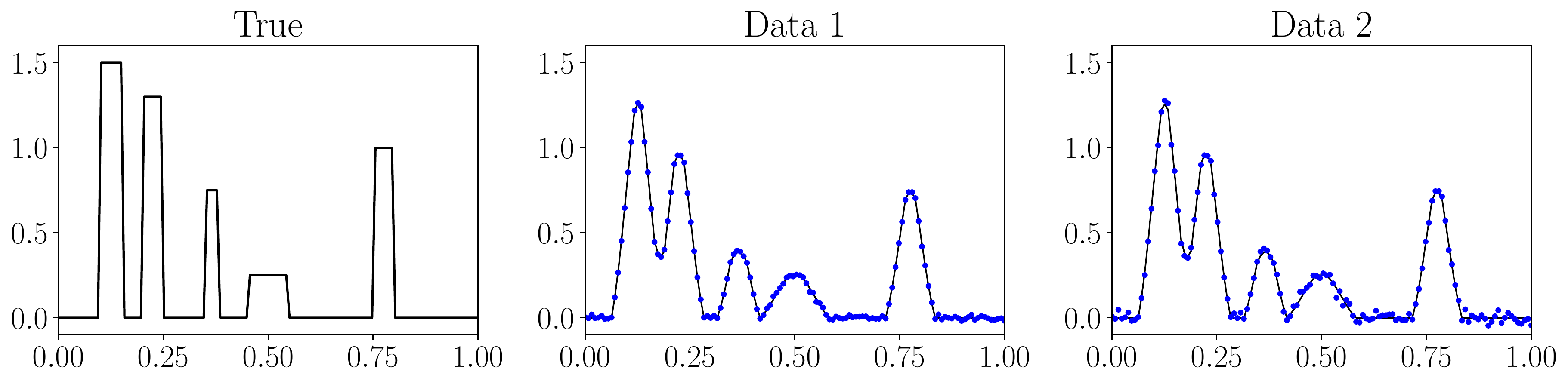}
\caption{Deconvolution example. Left: underlying true signal. Center: convolved and noisy data (dots) with $2\%$ noise. Right: convolved and noisy data with $5\%$ noise.}
\label{fig:ex_deconv_data}
\end{figure}

The inputs to the Gibbs sampler are studied/selected as follows: (i) The number of samples is $n_s=2\times 10^4$, in addition to a burn-in period of $n_b=2\times 10^3$. (ii) The scale parameter $\tau_0$ in the half-Cauchy hyperprior for $\tau$ is studied. (iii) The lag or thinning number $n_t$ is also analyzed. (iv) The influence of the number of standard and preconditioned CGLS iterations is studied as well.

We remark that for this example we can afford the factorization of the precision matrix $\widetilde{\mat{\Lambda}}$ in \cref{eq:postparams}. Hence, we can directly sample the Gaussian conditional $\pi_1$ in \cref{eq:cond1} without using CGLS. This allows us to obtain an exact sample from $\pi_1$ at each Gibbs iteration. This method is referred to as ``direct'' in the following studies and will be used to compare to the solution obtained by sampling $\pi_1$ with standard and preconditioned CGLS.

\paragraph{Influence of the scale parameter $\tau_0$} In this case, we use the low-noise data set, fix the lagging steps to $n_t=20$, and we directly sample the conditional $\pi_1$. Moreover, we fix the noise standard deviation to its true value. We perform a parameter study on $\tau_0$ using the values $\{(d_0/(d-d_0))\sigma_{\rm obs}^{\mathrm{true}}, \sigma_{\mathrm{obs}}^{\mathrm{true}}, 1\}$, where we assume the number of relevant variables is $d_0=10$, and thus, $\tau_0\in\{6.667\times 10^{-4}, 7.867\times 10^{-3}, 1\}$. The Gibbs sampler is applied for each value and we analyze the output of the Markov chain associated with the global parameter $\tau$, as well as the relative reconstruction error. For the point estimators, we also report the median since we observe that the posterior distribution of the $\tau$ parameter is right-skewed. The median is also used for the parameter $\ve{x}$ as it generates smaller reconstruction errors than the mean estimator.

The results are shown in \Cref{tab:tau0}. We observe that for the smallest value of $\tau_0$, the best reconstruction errors are obtained, however, the effective sample size $n_{\text{eff}}$ decreases. There are slight variations in the posterior mean and median of $\tau$ but overall the value remains in the same order of magnitude. In general, for this particular example, we do not observe significant changes in the results by modifying the parameter $\tau_0$. Nevertheless, from this simple study, we notice that the standard recommendation $\tau_0\approx\sigma_{\rm obs}^{\mathrm{true}}$ covers a middle ground in terms of the effective sample size and reconstruction error values.
\begin{table}[!ht]
\centering
\caption{Posterior mean and median of $\tau$, effective sampling size of the $\tau$-chain, and relative reconstruction error based on the mean and median estimators, for different values of the scale parameter $\tau_0$.}
\label{tab:tau0}
\begin{tabular}{c|ccc|cc}
\hline
$\tau_0$              & Mean & Median  & $n_{\text{eff}}$ & relerr (mean) & relerr (median) \\ \hline
$(d_0/(d-d_0))\sigma_{\rm obs}^{\mathrm{true}}$ &$6.218\times 10^{-3}$ & $5.675\times 10^{-3}$ & $4\,036$ & $1.413\times 10^{-2}$ & $1.099\times 10^{-2}$ \\
$\sigma_{\rm obs}^{\mathrm{true}}$ & $7.603\times 10^{-3}$& $7.059\times 10^{-3}$ & $4\,896$ & $1.571\times 10^{-2}$ & $1.216\times 10^{-2}$ \\
1                     & $9.096\times 10^{-3}$& $8.364\times 10^{-3}$ & $4\,426$ & $1.712\times 10^{-2}$ & $1.311\times 10^{-2}$ \\ \hline
\end{tabular}
\end{table}

\begin{remark}\label{rem:1}
From this study, we select the value of the scale parameter $\tau_0$ as the value of the noise standard deviation $\sigma_{\rm obs}$. Recall, however, that the true noise variance is unknown and hence we model it as a hyperparameter with associated conditional density $\pi_2$ in \cref{eq:cond_sigmaobs2}. Therefore, the value of $\tau_0^2$ is now equal to a realization of $\sigma_{\rm obs}^2$ at a given Gibbs iteration. Due to this modeling choice, the conditional density for the auxiliary parameter $\gamma$ in \cref{eq:cond_gamma} will now depend on $\sigma_{\rm obs}^2$, i.e., $\pi_5(\gamma\given \tau^2, \tau_0^2=\sigma_{\rm obs}^2)$, and hence, the sampling of $\gamma$ in line 13 of \Cref{algo:gibbs} can be modified accordingly. 

\end{remark}

\paragraph{Influence of the thinning $n_t$} As in the previous study, we use the small noise data set and we directly sample $\pi_1$. However, we now set the squared scale parameter $\tau_0^2$ equal to realizations of the noise variance random variable (cf. \Cref{rem:1}). This parameter study on the lag number $n_t$ uses the values $\{10, 20, 40, 80\}$. Once again, the Gibbs sampler is implemented with each value and we analyze the output of the Markov chain associated to the one-dimensional hyperparameters $\tau$ and $\sigma_{\rm obs}$, and also the reconstruction error.

The results are shown in \Cref{tab:nt}. The noise standard deviation $\sigma_{\rm obs}$ remains essentially unaffected by the thinning. For global parameter $\tau$, we observe that the IACT and consequently the number of effective samples improves with larger thinning. Particularly, we require a considerably high value of $n_t$ (in this case 80 or more) to obtain quasi-independent samples from $\tau$. Nevertheless, note that the mean and standard deviation estimators remain mostly unchanged by the thinning. We also compute the relative reconstruction error based on the median estimator of $\ve{x}$, with increasing thinning. The values for each parameter $n_t$ are $\text{relerr}=\{1.213,1.216,1.215,1.214\}\times 10^{-2}$, thus, the errors are approximately the same independent of the choice of $n_t$. Based on these results and to strike a balance between the computational cost and the quality of the posterior samples, the lag number for the thinning step can be selected between $20$ and $40$.
\begin{table}[!ht]
\centering
\caption{MCMC chain metrics and posterior statistics for the hyperparameters $\sigma_{\rm obs}$ and $\tau$, with increasing thinning values.}
\label{tab:nt}
\begin{tabular}{c|c|cccc}
\hline
Hyperparam. & $n_t$ & Mean & Std & IACT & $n_{\text{eff}}$ \\
\hline
\multirow{4}{*}{$\sigma_{\rm obs}$} & 10 & $7.945\times 10^{-3}$  & $5.585\times 10^{-4}$ & 1.06 & 18\,899 \\
& 20 & $7.947\times 10^{-3}$  & $5.559\times 10^{-4}$ & 1.03 & 19\,507 \\
& 40 & $7.949\times 10^{-3}$  & $5.584\times 10^{-4}$ & 1.03 & 19\,409 \\
& 80 & $7.948\times 10^{-3}$  & $5.614\times 10^{-4}$ & 1.05 & 19\,036 \\
\hline
\multirow{4}{*}{$\tau$} & 10 & $7.631\times 10^{-3}$ & $3.190\times 10^{-3}$ & 8.13 & 2\,460 \\
& 20 & $7.609\times 10^{-3}$ & $3.168\times 10^{-3}$ & 4.09 & 4\,888 \\
& 40 & $7.619\times 10^{-3}$ & $3.151\times 10^{-3}$ & 2.23 & 8\,986\\
 & 80 & $7.596\times 10^{-3}$ & $3.145\times 10^{-3}$ & 1.29 & 15\,399 \\
\hline
\end{tabular}
\end{table}

\paragraph{Standard and preconditioned CGLS} We use the settings obtained from the previous experiments, and now we select $n_t=40$. We run the Gibbs sampler with the draws from $\pi_1$ computed by standard and preconditioned CGLS\@. These are compared with the solution obtained by directly sampling from $\pi_1$. Both variants of CGLS are stopped at a given tolerance $\varepsilon_{\mathrm{cgls}}\in\{10^{-3},10^{-4}\}$. To compare the influence of the tolerance value in CGLS, we compute the relative reconstruction error in the solution.

\Cref{fig:cgls} shows the evolution of the CGLS iterations $n_{\rm cgls}$ with the number of samples obtained at every Gibbs iteration. These are shown for each predefined tolerance in the CGLS algorithm. Solid lines track the mean number of iterations and the vertical line marks the burn-in index for reference. We directly observe that preconditioned CGLS requires considerably less number of iterations per sample, compared to standard CGLS. 
\begin{figure}[!ht]
\centering
\includegraphics[width=0.95\textwidth]{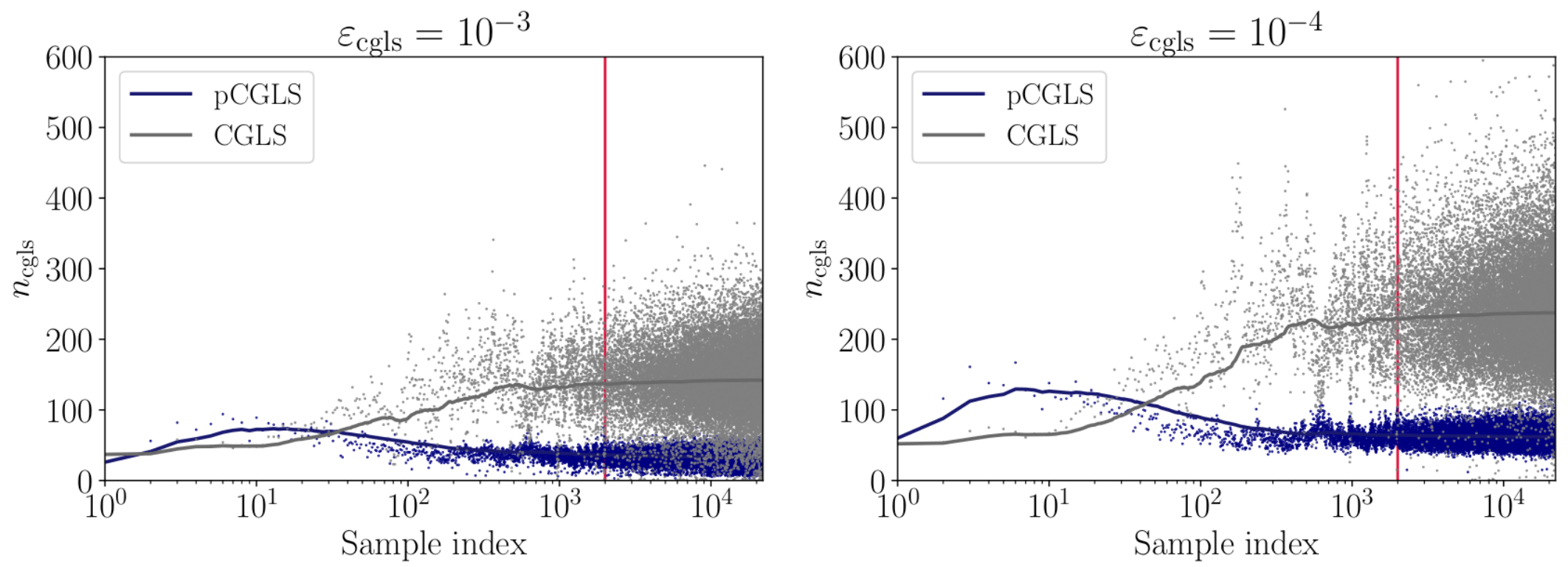}
\caption{Comparison of standard and preconditioned CGLS for different tolerances in the algorithm. The solid lines track the cumulative mean of the number of iterations. The solid vertical line marks the burn-in period.}
\label{fig:cgls}
\end{figure}

\Cref{tab:cgls} shows a detailed comparison of standard and preconditioned CGLS using both tolerance values. The relative reconstruction errors for the mean and median estimates of $\ve{x}$ are shown. To measure the quality of the results with respect to the posterior variability of $\ve{x}$, we compute the relative error of the standard deviation estimates; here we assume the standard deviation obtained by the direct method as true value. Moreover, the average number of CGLS/pCGLS iterations and the average effective sample size (across the $d$ components of the $\ve{x}$ chain) are also computed. We observe that with $\varepsilon_{\mathrm{cgls}}=10^{-4}$ CGLS and pCGLS give comparable results, and naturally pCGLS requires much less iterations. However, when $\varepsilon_{\mathrm{cgls}}= 10^{-3}$, the results of pCGLS deteriorate particularly in the estimation of the posterior variability. We see that the relative error in the posterior standard deviation is considerably large, while the errors in the mean and median estimates actually improve compared to the other cases. This indicates that, for this tolerance value, there is a concentration of the samples towards the mode, while missing the tails of the distribution. As a result, we require more stringent tolerances in pCGLS to guarantee convergence in order to obtain an exact sample from the Gaussian conditional $\pi_1$. In addition, we see that overall the samples of the parameter $\ve{x}$ are almost independent, given the average $n_{\rm eff}$ values are close to $n_s$.
\begin{table}[!ht]
\centering
\caption{Comparison of standard and preconditioned CGLS for different CGLS tolerances. The direct method that samples exactly from $\pi_1$ is also shown.}
\label{tab:cgls}
\begin{tabular}{ll|ccccc}
\hline
                    &    & relerr (mean) & relerr (median) & relerr (std) & Average $n_{\text{cgls}}$ & Average $n_{\rm eff}$ \\
\hline
\multicolumn{2}{c|}{Direct}  &  $1.56\times 10^{-2}$ & $1.21\times 10^{-2}$    &    --     &  --  & 19\,386   \\
\hline
\multirow{2}{*}{$\varepsilon_{\mathrm{cgls}}=10^{-3}$} & CGLS  &  $1.52\times 10^{-2}$    &  $1.20\times 10^{-2}$    &     $5.81\times 10^{-2}$    &  143  &  19\,291  \\
                    & pCGLS &  $0.89\times 10^{-2}$  & $0.69\times 10^{-2}$  &   $34.5\times 10^{-2}$ &  33  & 18\,857   \\
\hline
\multirow{2}{*}{$\varepsilon_{\mathrm{cgls}}=10^{-4}$} & CGLS  & $1.55\times 10^{-2}$  & $1.21\times 10^{-2}$ & $1.99\times 10^{-2}$ & 238 & 19\,384  \\
                    & pCGLS & $1.54\times 10^{-2}$  & $1.19\times 10^{-2}$ &  $1.78\times 10^{-2}$ & 62 & 19\,369 \\
\hline
\end{tabular}
\end{table}

\paragraph{Solution for different noise levels} \yq{Now we use both data sets.} The solutions are computed using preconditioned CGLS with $\varepsilon_{\mathrm{cgls}}=10^{-4}$. First we show in \Cref{fig:sigobs_tau_low} posterior statistics for the hyperparameters $\sigma_{\mathrm{obs}}$ and $\tau$ for the low noise data set. In particular, we plot the histogram of the posterior samples and generated chain, as well as, the ergodic mean and the sample autocorrelation. We observe that the posterior for the global parameter $\tau$ is right-skewed, while the posterior of $\sigma_{\rm obs}$ is almost symmetric. The posterior samples computed by the Gibbs sampler have significantly small correlation and we also see that the posterior chain has converged and it is well-mixed. We obtain similar results for the large noise data set and thus we omit the figure.
\begin{figure}[!ht]
\centering
\includegraphics[width=0.99\textwidth]{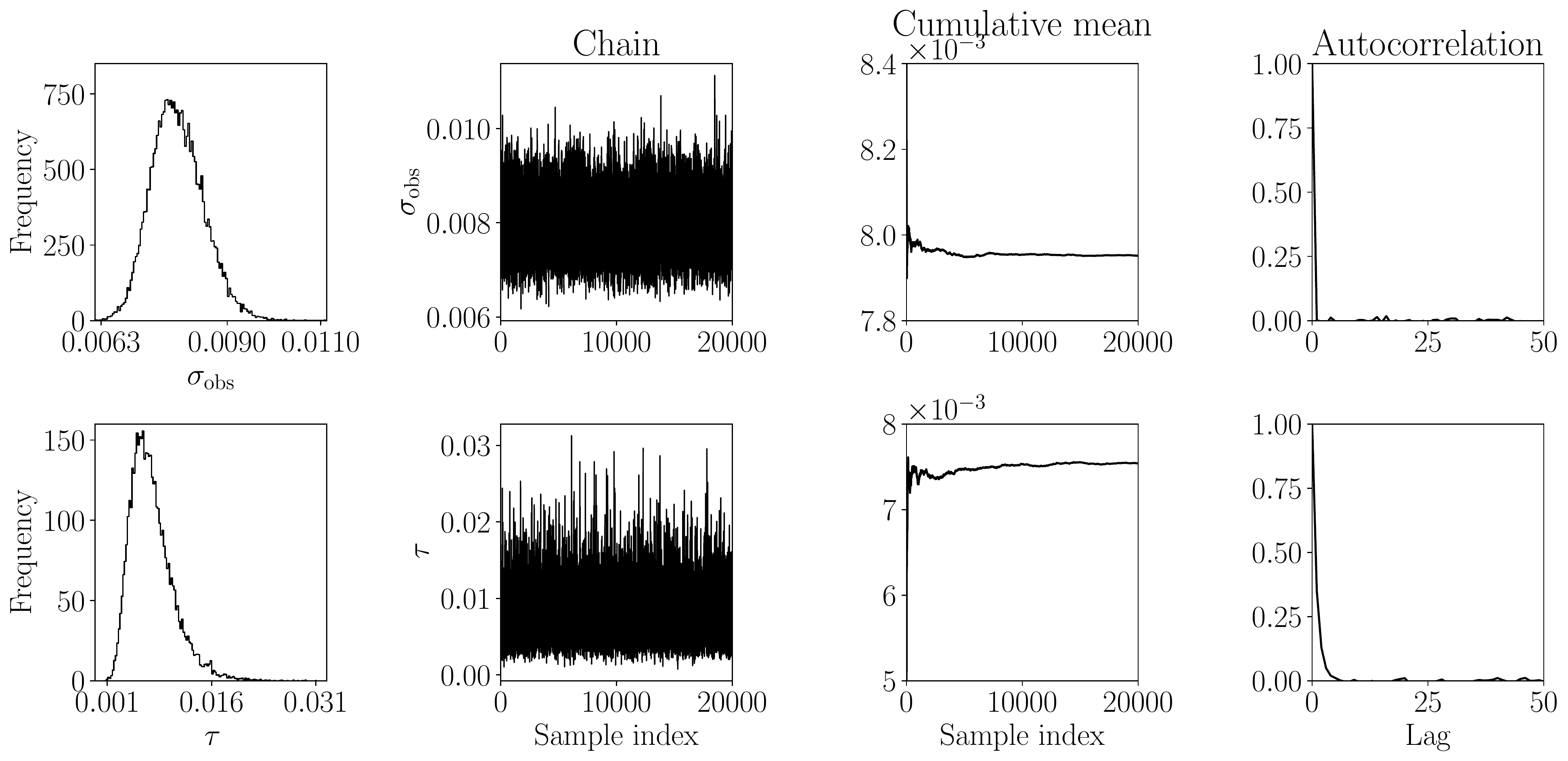}
\caption{Posterior of $\sigma_{\rm obs}$ (first row) and $\tau$ (second row) using the low noise data set. Histogram, chain, cumulative mean and sample autocorrelation.}
\label{fig:sigobs_tau_low}
\end{figure}

\Cref{fig:w_low_med} shows the posterior mean and 95\% credible interval for the local weight parameter $\ve{w}$. Note that the local parameter captures the locations where the signal edges are placed. The uncertainty is large at those particular locations and essentially zero in the rest of the domain. We observe that as the noise increases, the magnitude of the weights decreases and the point estimators become less sharp. Nevertheless, the posterior is still able to capture the edges correctly.
\begin{figure}[!ht]
\centering
\includegraphics[width=0.95\textwidth]{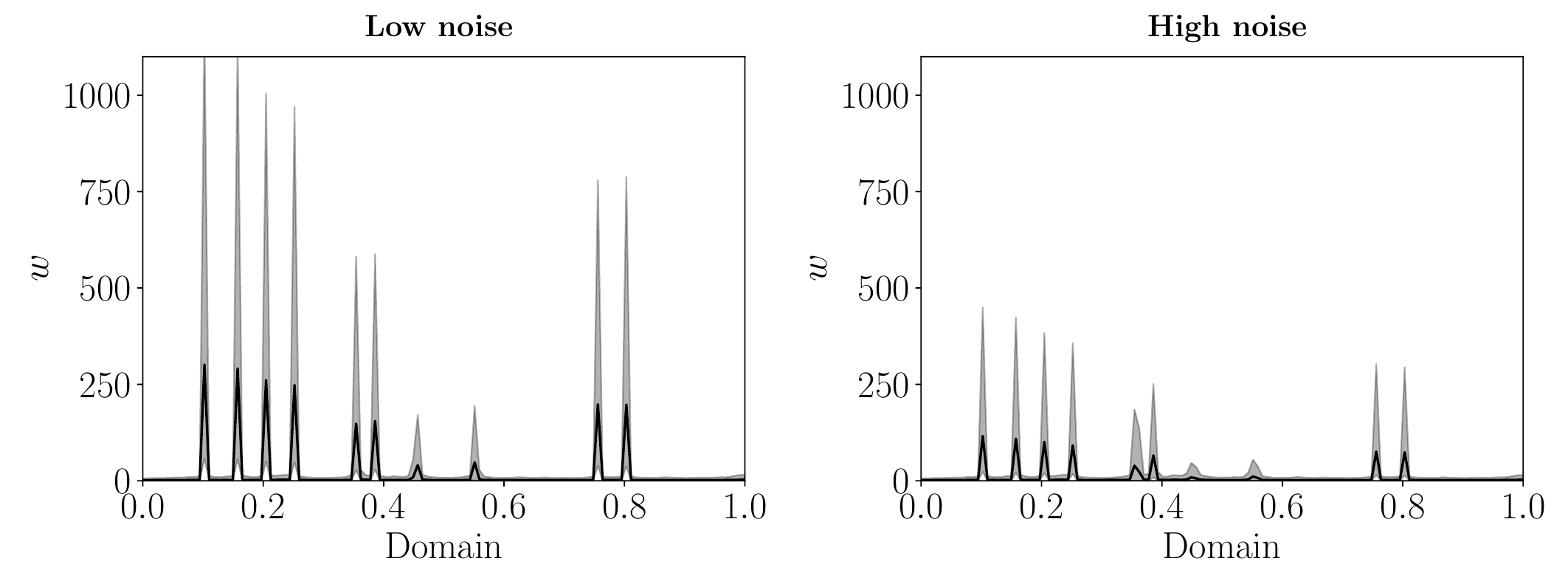}
\caption{Posterior statistics for $\ve{w}$ using the low noise (left) and high noise (right) data sets. Posterior mean (solid line) and 95\% credible interval (shaded area).}
\label{fig:w_low_med}
\end{figure}

Finally, the main posterior statistics for the target parameter $\ve{x}$ are shown in \Cref{fig:x_low_med} together with the underlying true signal. Similar to the local weights, we plot the posterior mean and 95\% credible interval. The relative reconstruction errors based on the mean, for the low and high noise cases, are $1.54\times 10^{-2}$ and $6.63\times 10^{-2}$, respectively. We compare the results based on the horseshoe prior to the method in \cite{uribe_et_al_2021} which is based on a Laplace Markov random field prior. For the latter, we use the same number of samples, burn-in and thinning. In the case of the Laplace Markov random field prior, we obtain the reconstruction errors $5.36\times 10^{-2}$ and $9.27\times 10^{-2}$, for each data set. \yq{We see that our method not only generates sharper solutions but also reduces the posterior uncertainty.}
\begin{figure}[!ht]
\centering
\includegraphics[width=0.99\textwidth]{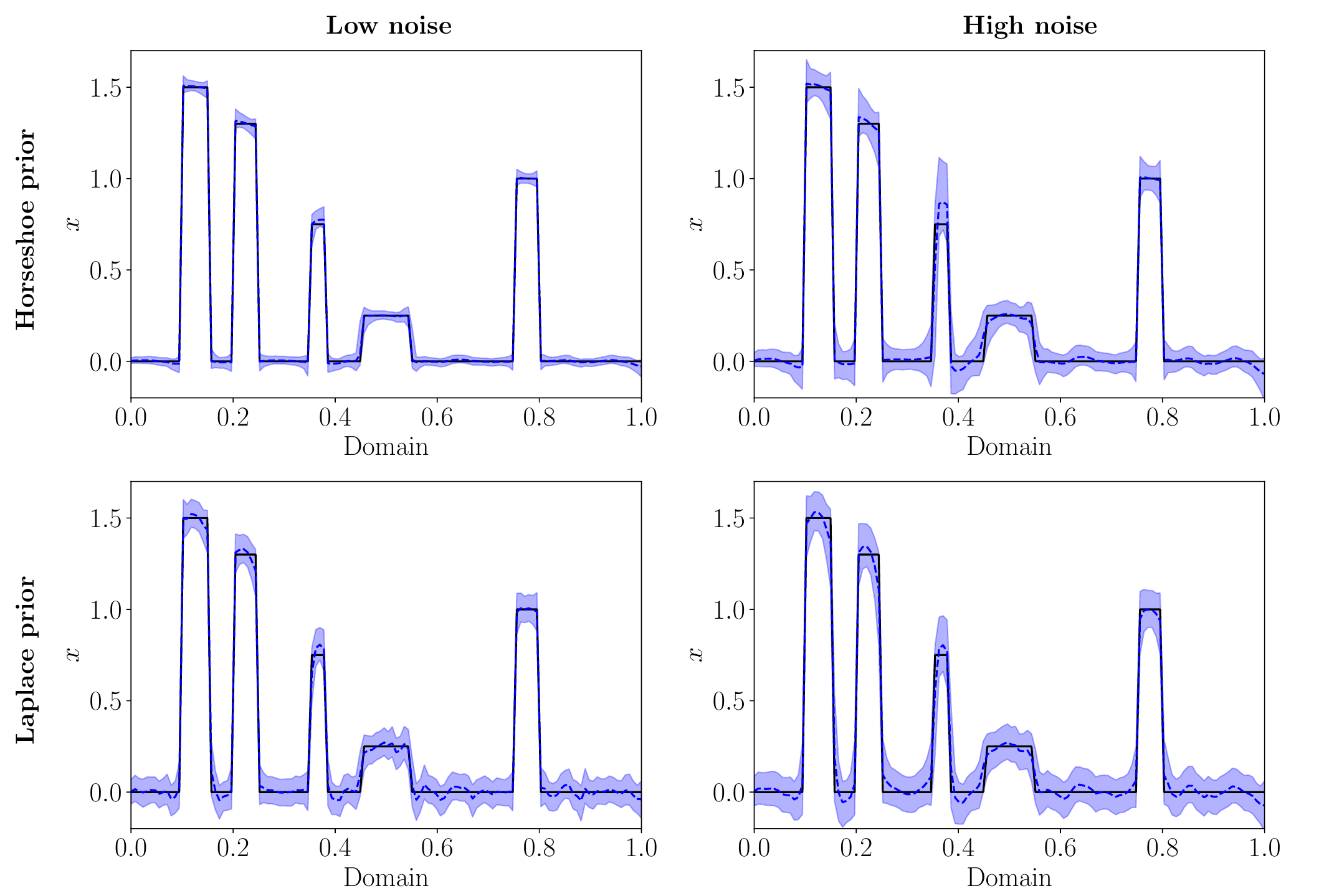}
\caption{\yq{Comparison of posterior statistics for the target parameter $\ve{x}$ using the horseshoe prior (first row) and the Laplace prior (second row).} The results are shown using the low noise (left) and high noise (right) data sets. True solution (solid line), posterior mean (dashed line) and 95\% credible interval (shaded area).}
\label{fig:x_low_med}
\end{figure}

\subsection{Two-dimensional examples}
In this section, we study a couple of linear inverse problems to test the horseshoe prior in two-dimensional applications. In both examples, the inputs to the Gibbs sampler are selected as follows: (i) The number of samples is $n_s=2\times 10^4$, in addition to a burn-in period of $n_b=4\times 10^3$. (ii) The scale parameter $\tau_0$ is defined from the noise standard deviation as mentioned in \Cref{rem:1}. (iii) The thinning number is set to $n_t=20$. (iv) In two-dimensional applications the computation of the QR decomposition in \cref{eq:QR} required in pCGLS dominates the computational cost. Instead, we design the experiments such that the direct method to sample $\pi_1$ can be utilized.

\subsubsection{Image deblurring}
We consider an image deblurring inverse problem where we seek to recover an original sharp image from a blurred one by using a mathematical model of the blurring process. This is analogous to one-dimensional deconvolution but now in two dimensions. In this case, we assume that the blurring of the columns in the image is independent of the blurring of the rows. Therefore, there exist two matrices $\mat{A}_{\rm c}\in\mathbb{R}^{N\times N}$ and $\mat{A}_{\rm r}\in\mathbb{R}^{N\times N}$, such that we can express the forward operator as $\mat{A}_{\rm c}\mat{X}\mat{A}_{\rm r}^\tran$ with an image $\mat{X}\in\mathbb{R}^{N\times N}$. Here, we set $\mat{A}_{\rm c}$ and $\mat{A}_{\rm r}$ as in Challenge 1 from \cite{hansen_et_al_2006}; in MATLAB notation these are:
\begin{align}
&\texttt{c = zeros(N,1);  c(1:5) = [5:-1:1]'/15;  Ac = toeplitz(c)}\nonumber\\
&\texttt{r = zeros(N,1);  r(1:10) = [5:-.5:.5]'/15;  Ar = toeplitz(c,r)}.\nonumber
\end{align}

Since we typically work with a vectorized version of the image, we can define the system matrix as $\mat{A} = (\mat{A}_{\rm r}\otimes \mat{A}_{\rm c})\in\mathbbm{R}^{d\times d}$, such that the linear forward operator becomes $\mat{A}\ve{x}$. In this example, the image size is $32$-by-$32$, i.e., $N=32$ and $d=N^2=1\,024$. Moreover, we set the Gaussian noise level to $1\%$. The underlying true image and its blurred and noisy version are shown in \Cref{fig:ex_deblur_data}. 
\begin{figure}[!ht]
\centering
\includegraphics[width=0.75\textwidth]{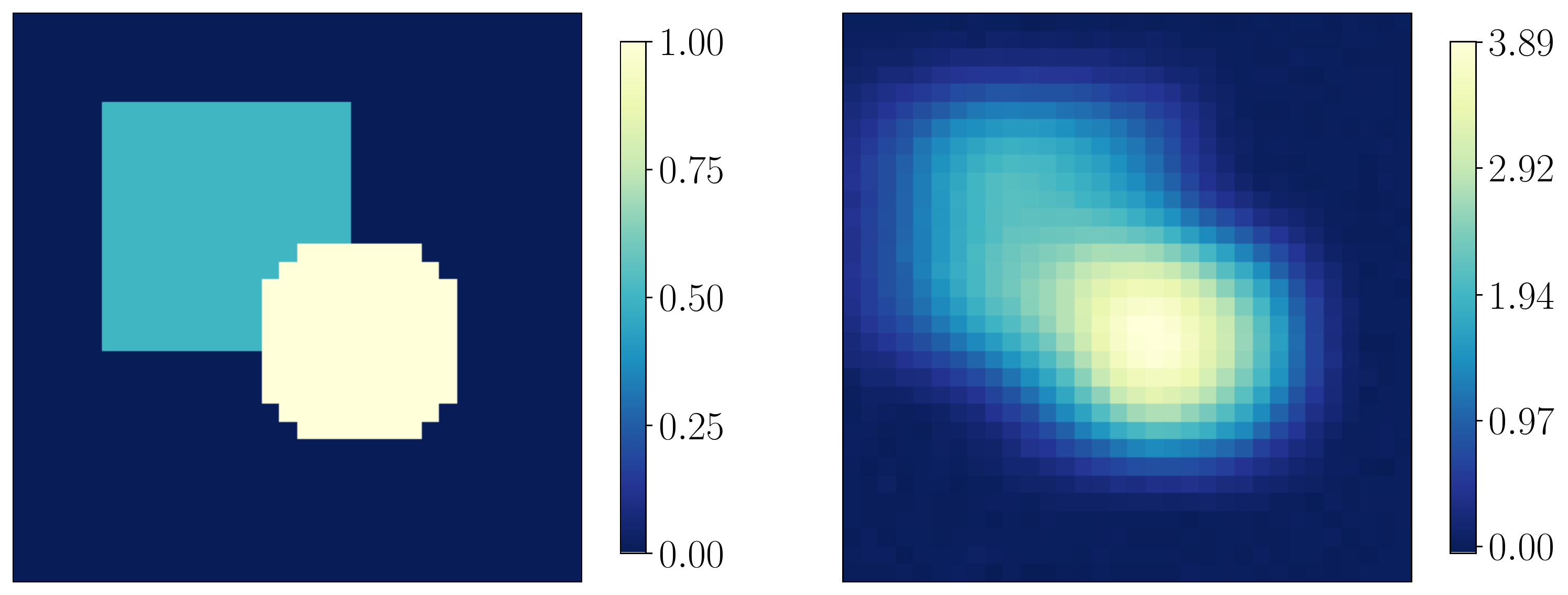}
\caption{Image deblurring example. Left: underlying true image. Right: blurred noisy image.}
\label{fig:ex_deblur_data}
\end{figure}

The posterior solution for the hyperparameters $\sigma_{\mathrm{obs}}$ and $\tau$ is shown in \Cref{fig:sigobs_tau_low_deblur}. As in the previous example, we plot the histogram of the posterior samples, the Markov chain, the ergodic mean and the sample autocorrelation. The magnitude of both hyperparameters is very small, especially the global standard deviation $\tau$ that tends to shrink parameters to zero. Furthermore, we observe that the posterior samples computed by the Gibbs sampler have small correlation and that the chain has converged and it is well-mixed.
\begin{figure}[!ht]
\centering
\includegraphics[width=0.99\textwidth]{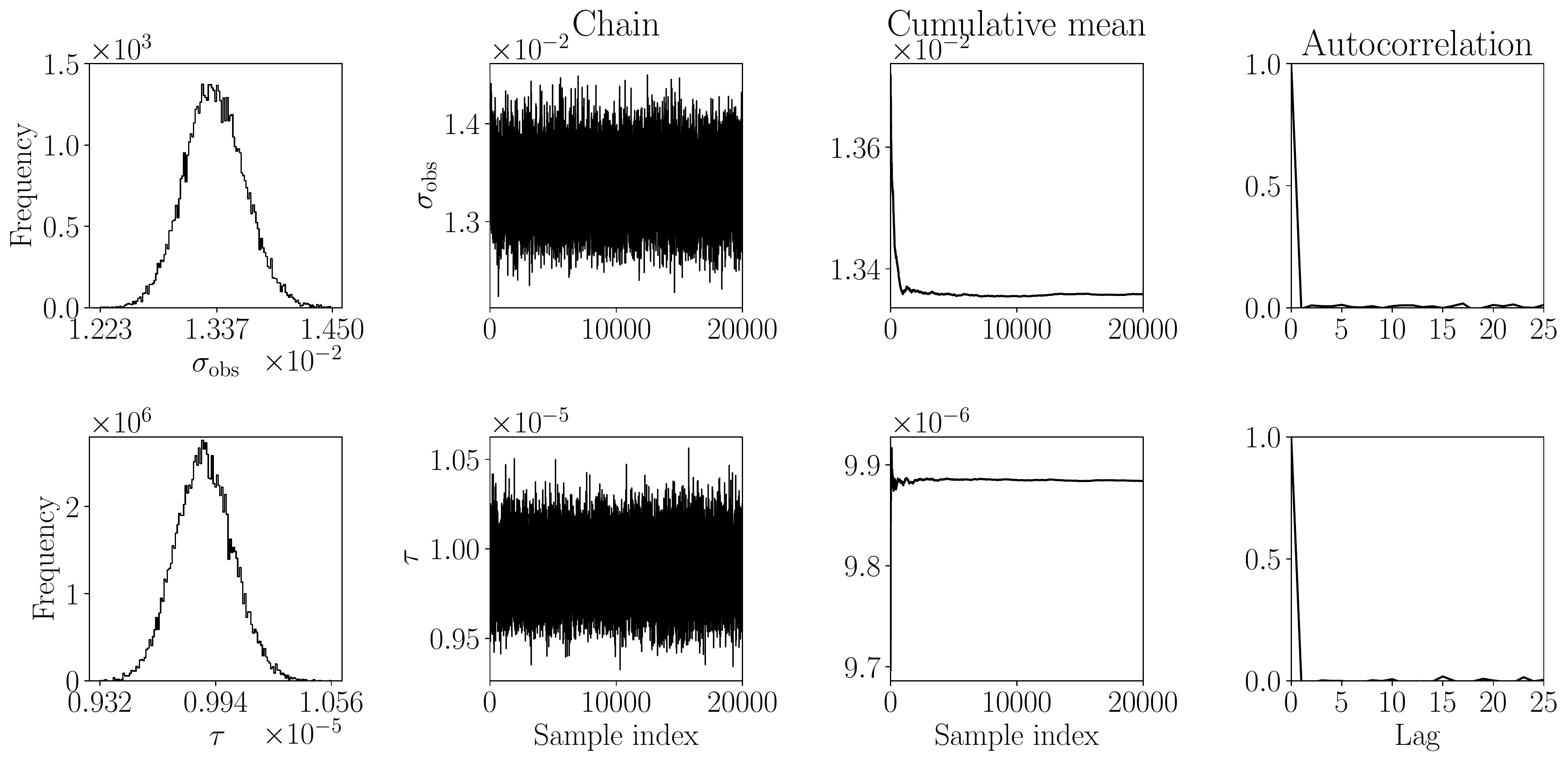}
\caption{Posterior of $\sigma_{\rm obs}$ (first row) and $\tau$ (second row). Histogram, chain, cumulative mean and sample autocorrelation. Note the sharp decay in the autocorrelation after first lag, an indication of quasi-independent samples.}
\label{fig:sigobs_tau_low_deblur}
\end{figure}

\Cref{fig:w_deblur} shows the posterior mean and standard deviation for the local weight parameter $\ve{w}$. Note that the local parameter captures the locations where the signal edges are placed. Particularly, the local parameter $\ve{w}^{(1)}$ captures the locations where the image vertical edges are located, while the weight $\ve{w}^{(2)}$ identifies the horizontal edges. Similar as in the one-dimensional example, the weights and their uncertainty are large at the edge locations in order to escape shrinkage, and essentially zero in the rest of the domain. 
\begin{figure}[!ht]
\centering
\includegraphics[width=0.77\textwidth]{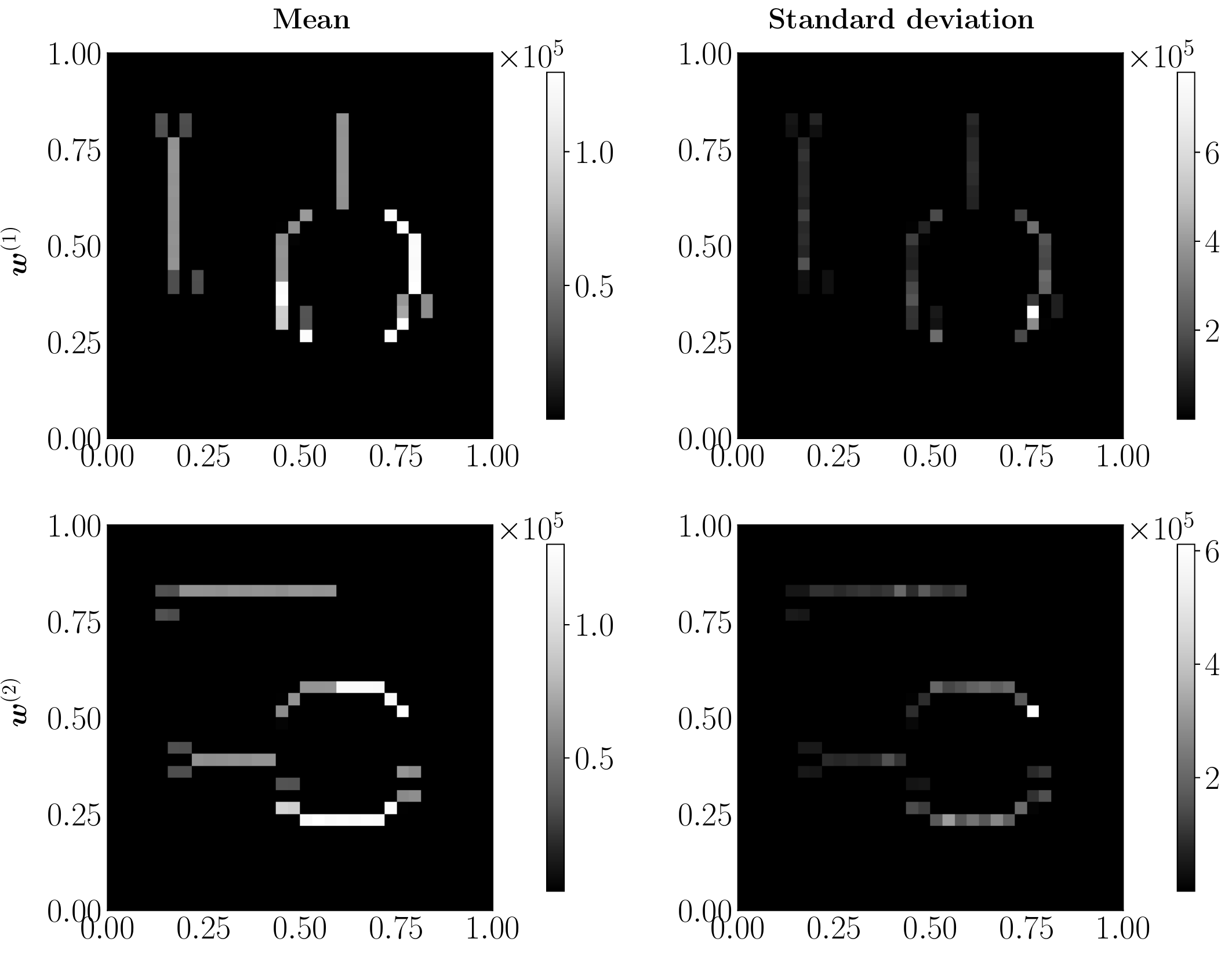}
\caption{Posterior statistics for the local weights $\ve{w}^{(1)}$ (first row) and $\ve{w}^{(2)}$ (second row). Posterior mean (left) and standard deviation (right).}
\label{fig:w_deblur}
\end{figure}

Posterior statistics for the image represented by $\ve{x}$ are shown in \Cref{fig:x_deblur}. We compare the posterior mean and standard deviation obtained by using the horseshoe prior with those from the Laplace Markov random field prior \cite{uribe_et_al_2021}. We show the posterior mean with the same intensity range in both results for the sake of comparison, however, there are no negative values in the horseshoe case. The relative reconstruction error based on the mean are $0.098$ and $0.065$, respectively. Despite the relative error is slightly larger for the horseshoe prior in this example, we clearly see that the mean of the posterior based on the horseshoe prior is sharper and has smaller posterior uncertainty. Nevertheless, with the horseshoe prior the standard deviation is particularly large at the locations that are not identified properly, while with the Laplace prior the larger values of standard deviation are located at the edges.
\begin{figure}[!ht]
\centering
\includegraphics[width=0.76\textwidth]{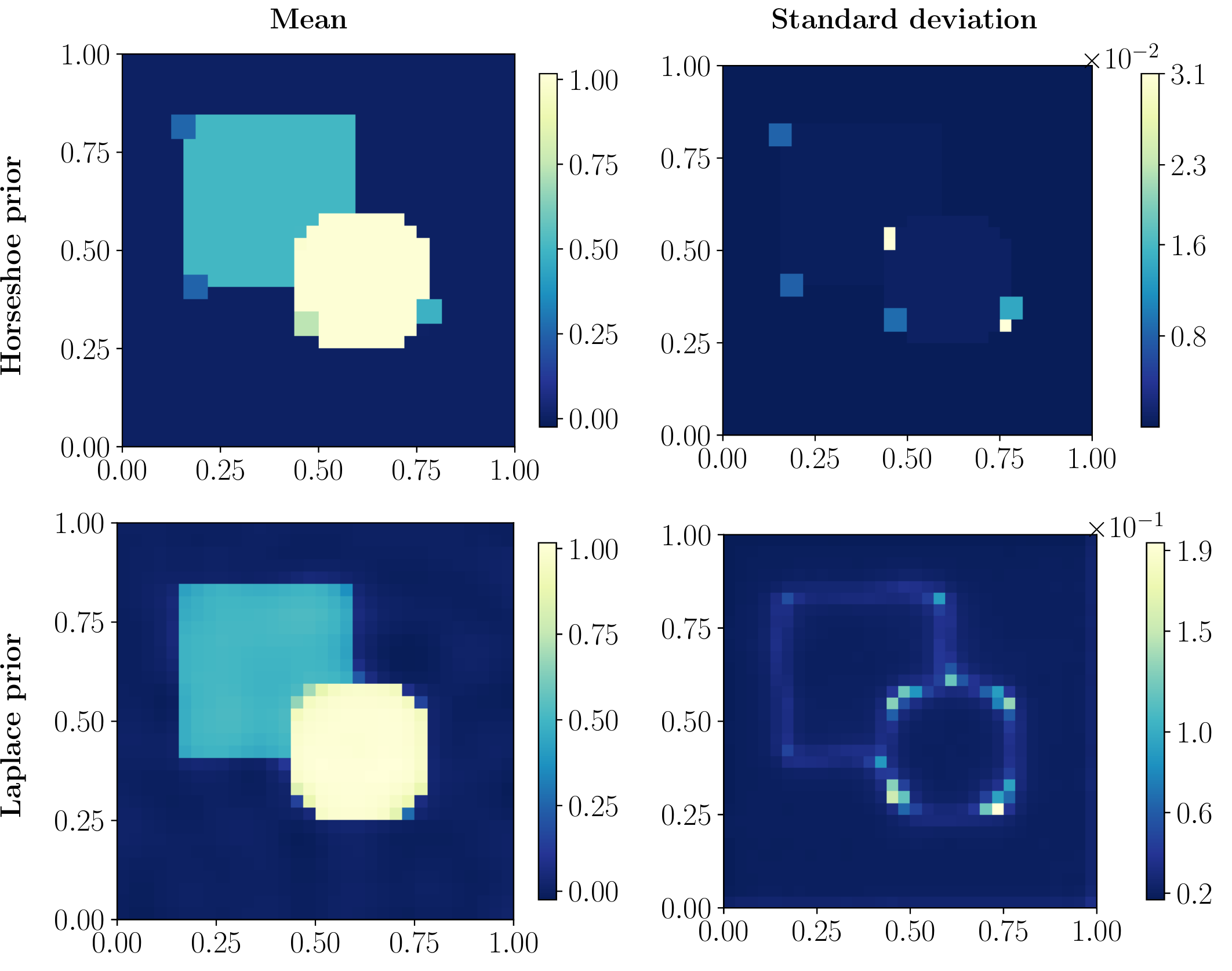}
\caption{Comparison of posterior statistics for the target parameter $\ve{x}$ using the horseshoe prior (first row) and the Laplace prior (second row). Posterior mean and standard deviation.}
\label{fig:x_deblur}
\end{figure}

Finally, notice that for the posterior based on the horseshoe prior most of the error contribution is associated to the misrepresentation of a few corners of the objects in the image. To illustrate this we plot a few posterior samples in \Cref{fig:x_deblur_sampls}. In some cases, the samples detect the corner pixels with interior intensity but others with the outside intensity (see, e.g., bottom right corner of the ``round'' object). We observe that with the horseshoe prior the edges are always sharp, but those arising from the Laplace prior are in general more noisy. This is related to the fact that the horseshoe prior is more heavy tailed than the Laplace prior.
\begin{figure}[!ht]
\centering
\includegraphics[width=0.98\textwidth]{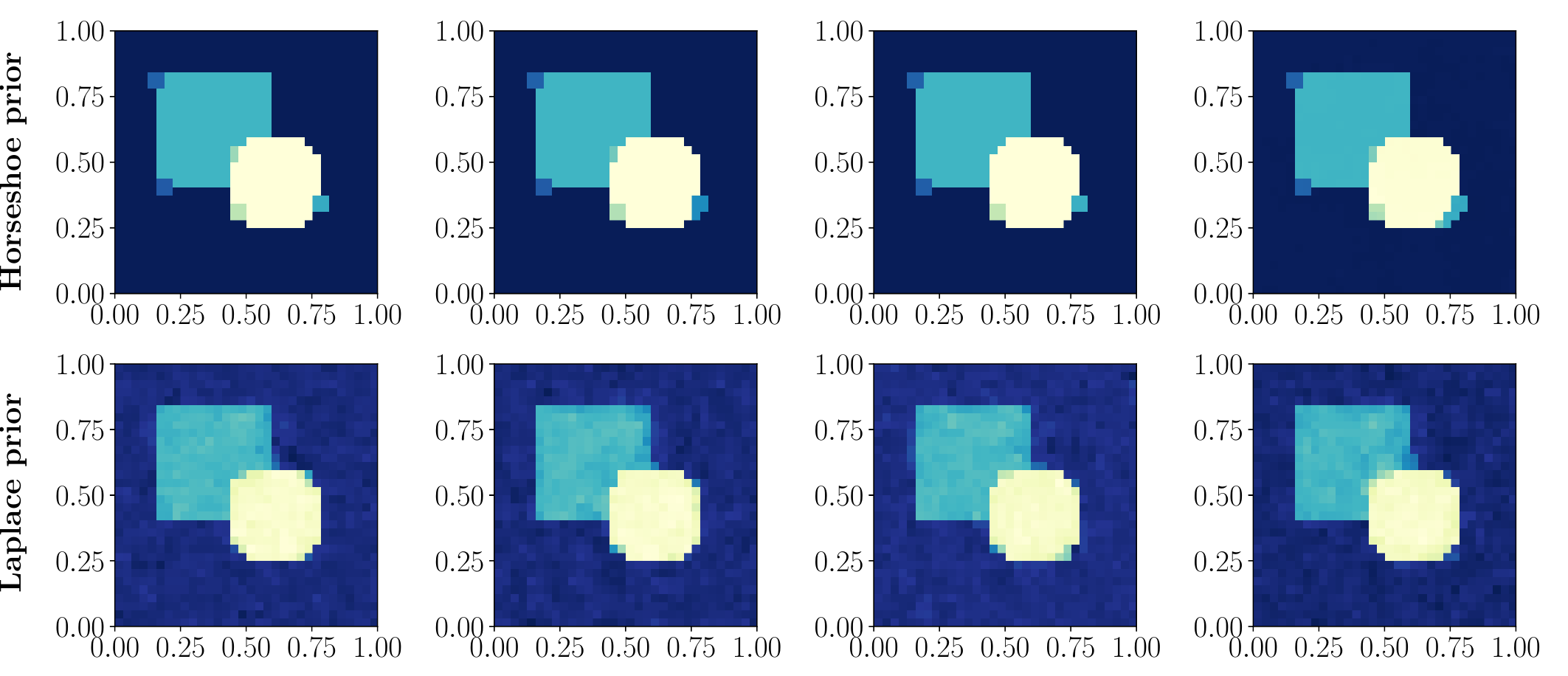}
\caption{Comparison between a few posterior samples generated from the horseshoe prior (first row) and Laplace prior (second row).}
\label{fig:x_deblur_sampls}
\end{figure}

\subsubsection{X-ray CT reconstruction}
The objective of 2D CT reconstruction is to estimate an image of the cross-section of an object using projection data that represents the intensity loss or attenuation of a beam of X-rays as they pass through the object. 

The discretized CT reconstruction problem takes the form given in \eqref{eq:inv_prob}, where $\ve{x}^{\mathrm{true}}\in \mathbbm{R}^d$ is the vector of attenuation coefficients, which corresponds to an $N$-by-$N$ matrix with $d=N^2$, and $\ve{y}\in\mathbbm{R}^m$ denotes the measurement data. The system matrix $\mat{A}\in\mathbbm{R}^{m\times d}$ is from the discretized Radon transform, and the number of rows $m$ is the product of the number of detector elements $p$ and the number of projection angles $q$ (see, e.g., \cite[Ch.9]{hansen_et_al_2021} for the details). In this context, $\ve{y}$ is referred to as projection data that is typically expressed as a 2D array known as sinogram. 

In particular, we consider the reconstruction of a ``{grains}'' phantom generated with the \texttt{phantomgallery} function from \cite{hansen_and_jorgensen_2018}. The projection geometry and data acquisition process comes from a \emph{fan beam} configuration \cite[p.53]{hansen_et_al_2021}, whose parameters are defined as follows (in arbitrary units): the distance from the X-ray source to the origin is $3N$, the distance from the origin to the detector is $N$, and the number of detector elements is $p=N$. The discretized domain size is set to $N=64$ (thus $d=4\,096$), and we use $q=32$ equidistantly distributed projection angles in $[0^\circ, 360^\circ]$. Furthermore, the noise standard deviation is set as ${\sigma}_{\mathrm{obs}}= 0.01 (\norm{\mat{A} \ve{x}^{\mathrm{true}}}_2/ \sqrt{m})$ with $m=pq=2\,048$. \Cref{fig:grains_true_data} shows the true image together with the sinogram data.
\begin{figure}[!ht]
\centering
\includegraphics[width=0.75\textwidth]{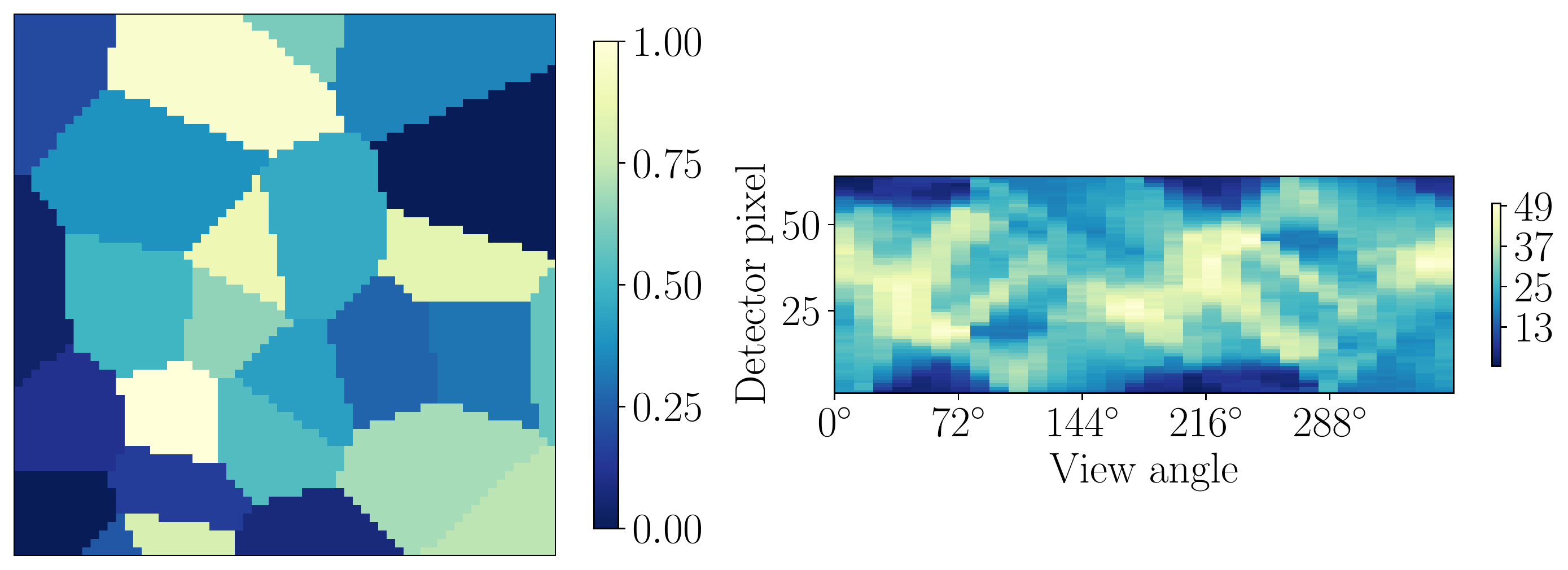}
\caption{Grains phantoms. Left: underlying true image. Right: noisy sinogram data.}
\label{fig:grains_true_data}
\end{figure}

For the hyperparameters $\sigma_{\mathrm{obs}}$ and $\tau$ we obtain a similar behavior to the previous examples (hence we omit the plot). The posterior mean and standard deviation of $\sigma_{\mathrm{obs}}$ and $\tau$ are $0.349\pm 0.006$ and $4.94\times 10^{-6}\pm 3.84\times 10^{-8}$, respectively. The IACTs for both hyperparameters are close to 1, and thus, we also obtain quasi-independent samples using the proposed Gibbs sampler in this example.

The posterior mean and standard deviation for the local weight parameter $\ve{w}$ are shown in \Cref{fig:CT2D_w}. Similarly to the deblurring example, we see that $\ve{w}^{(1)}$ captures the locations of vertical edges and $\ve{w}^{(2)}$ finds the horizontal edges. In this case, it is more noticeable the influence of the assumed zero boundary conditions on the image; see the leftmost column in $\ve{w}^{(1)}$ and the topmost row in $\ve{w}^{(2)}$. We also observe that some of the edges are missed, especially where there is no significant differences in the attenuation coefficients of the grains (see, e.g., bottom right area in the posterior means).
\begin{figure}[!ht]
\centering
\includegraphics[width=0.73\textwidth]{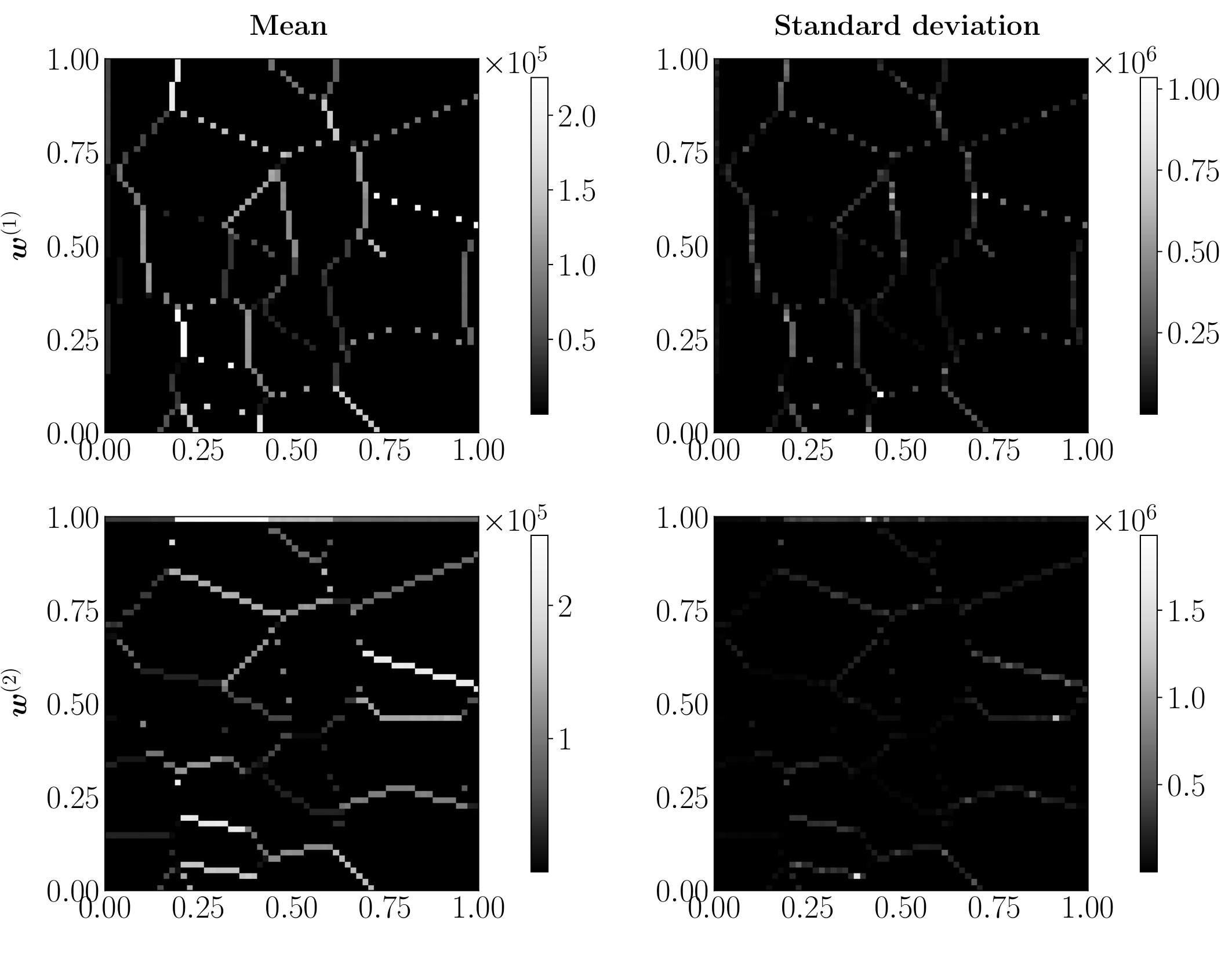}
\caption{Posterior statistics for the local weights $\ve{w}^{(1)}$ (first row) and $\ve{w}^{(2)}$ (second row). Posterior mean (left) and standard deviation (right).}
\label{fig:CT2D_w}
\end{figure}
\begin{figure}[!ht]
\centering
\includegraphics[width=0.73\textwidth]{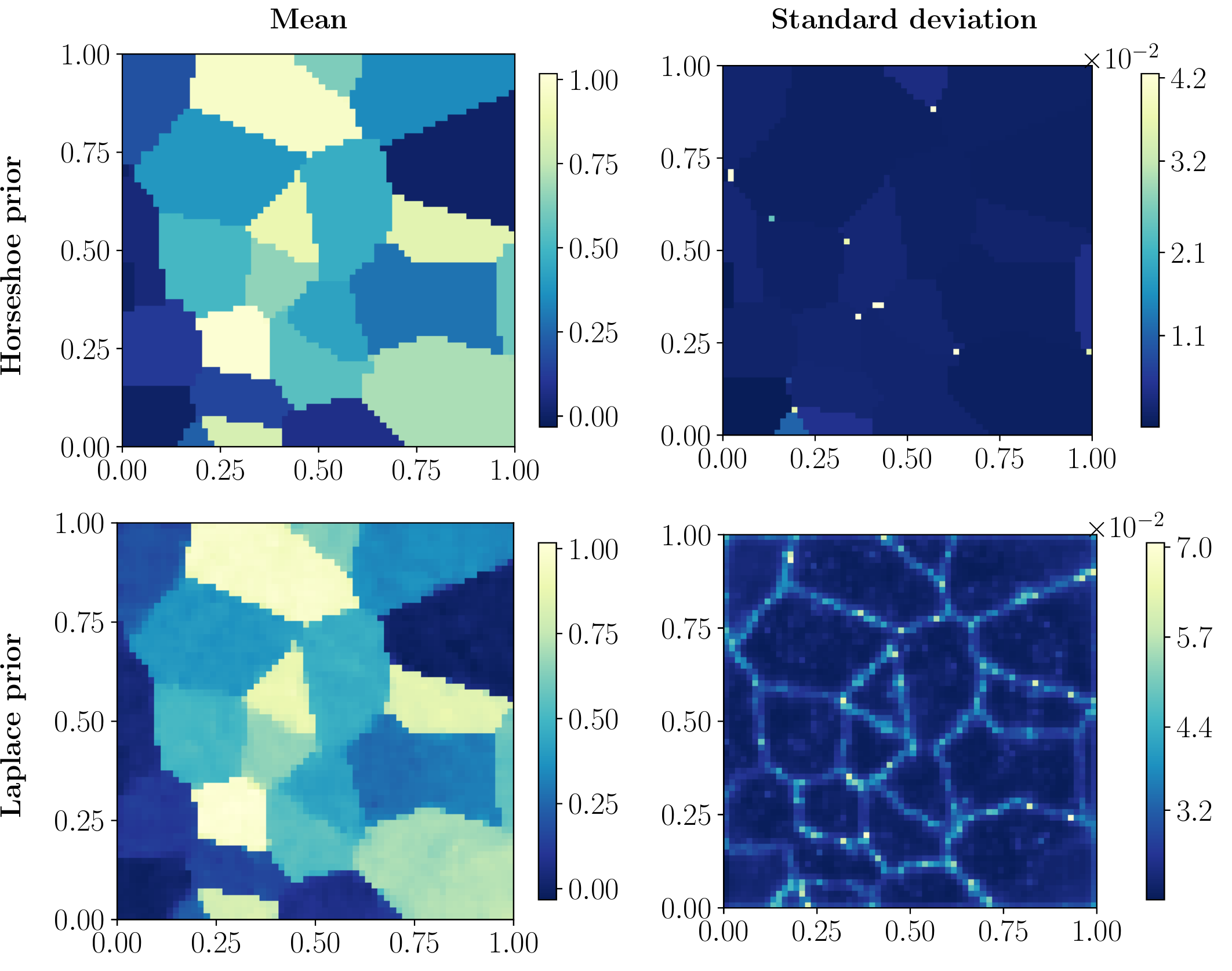}
\caption{Comparison of posterior statistics for the target parameter $\ve{x}$ using the horseshoe prior (first row) and the Laplace prior (second row). Posterior mean and standard deviation.}
\label{fig:CT2D_x}
\end{figure}

The main posterior statistics for the parameter $\ve{x}$ are shown in \Cref{fig:CT2D_x}. We plot the posterior mean and standard deviation, and compare them with the results from using the Laplace Markov random field prior. Again, we show the posterior mean with the same intensity range for the sake of comparison, but no negative values are obtained in the horseshoe case. Using the posterior mean as point estimate, the relative reconstruction error based on the mean are $0.045$ for the horseshoe prior and $0.052$ for the Laplace prior. Similarly to the other examples, it is clear that the horseshoe prior allows us to obtain a sharper reconstructions with smaller posterior uncertainty compared to the Laplace prior. Despite some of the grain features are missed in the reconstruction, we observe that the standard deviation values reflect higher uncertainty at those particular missing locations. Moreover, note there are a few pixels at which the uncertainty is larger compared to the others; as in the deblurring example, these tend to be located at the corners of the grains.

\section{Summary and conclusions}\label{sec:conclusions}
We developed a computational framework for solving linear Bayesian inverse problems where preserving edges or sharp features of the solution is required. Our approach is based on the hierarchical horseshoe prior which combines a conditionally Gaussian prior on pairwise parameter increments with two half-Cauchy hyperpriors on global and local variance hyperparameters. The global hyperparameter reduces small increments towards zero, while the local hyperparameter allows some sufficiently large increments to escape shrinkage, thereby detecting the edges.

Due to the hierarchical structure and heavy-tailed distributions involved in the prior, sampling of the posterior distribution is cumbersome. We employ a scale mixture representation of the half-Cauchy hyperpriors to obtain a formulation with distributions of simpler form. Thereafter, sampling of the posterior is performed via the Gibbs sampler where the conditional densities are derived in closed form by exploiting conjugate relations. In particular, the conditional density for the target parameter is high-dimensional Gaussian and we formulate the task of sampling as a least-squares problem that is solved efficiently via CGLS with preconditioner given by the prior precision matrix. We point out that despite our method is tailored for linear inverse problems, it can also be extended to nonlinear ones by sampling the conditional densities for the target parameter and the noise variance hyperparameter with suitable MCMC samplers. 

The numerical experiments show that our approach based on the horseshoe prior enables the computation of sharp point estimates of the posterior, while reducing the posterior uncertainty. This compared with more traditional prior models based on Laplace Markov random fields. Furthermore, in one-dimensional applications the computational cost of the algorithm is significantly low since the preconditioner for CGLS is obtained as a byproduct of the formulation. However, in large-scale two-dimensional problems our approach becomes prohibitively slow since the preconditioner is expensive to compute. 

Finally, some ideas to extend/improve our methodology are as follows:
\begin{itemize}
\item[(i)] Further work is needed to reduce the computational cost of sampling the Gaussian conditional in large-scale inverse problems. Methods from linear algebra such as inverse square root and matrix splitting approximations can be useful (see, e.g., \cite{vono_et_al_2022}), or MCMC methods that exploit low-dimensional structure \cite{martin_et_al_2012}, as well as, finding easy-to-compute preconditioners for CGLS. 
\item[(ii)] Application of the regularized horseshoe prior is a reasonable extension of our framework to make the posterior inference more robust (see, e.g., \cite{nishimura_and_suchard_2022}). The idea is to control potential issues arising from the heavy-tails of shrinkage priors. 
\item[(iii)] The Bayesian inverse problem can be re-parameterized in terms of the increments (see, e.g., \cite{calvetti_et_al_2020b}), in this case, the prior precision matrix is equal to the weight matrix. Despite this transformation complicates the likelihood function, this can prove useful when computing the preconditioner required in pCGLS since the precision matrix is now diagonal. 
\item[(iv)] In two-dimensional settings, the local weights which are associated to each coordinate direction, can be reduced to the have the same size of the target parameter. The idea is to identify redundant edges by adding compatibility conditions (see, e.g., \cite{calvetti_et_al_2020b}). 
\item[(v)] The number of hyperparameters can be reduced using empirical Bayes or maximum marginal likelihood methods. The idea is to fix the hyperparameter to the value that maximizes the model evidence. This strategy is used for example in sparse Bayesian learning \cite{tipping_2001}.
\end{itemize}

\section*{Acknowledgements}\label{sec:acknowledgements}
This work was supported by a Villum Investigator grant (no.\ 25893) from The Villum Foundation.

\addcontentsline{toc}{section}{References}
\bibliographystyle{model1-num-names}
\bibliography{bibfile.bib}

\end{document}